\documentclass[11pt]{article}

\usepackage{amsmath,amsfonts,amssymb,amsbsy,textcomp}
\usepackage{graphicx,color}
\usepackage{hyperref}
\usepackage{graphics}
\usepackage{epsfig}
\usepackage{verbatim}
\usepackage{amsfonts}

\usepackage{epsf}
\def\ssl#1{\rlap{\hbox{$\mskip 3 mu /$}}#1}
\def\be{\begin{equation}}
\def\ee{\end{equation}}
\def\bc{\begin{center}}
\def\ec{\end{center}}
\def\ba{{\bar{a}}}
\def\bR{{\mathbb{R}}}
\def\bD{{\mathbb{D}}}

\def\by{{\mathbf{y}}}
\def\bbZ{{\mathbb{Z}}}
\def\cB{{\mathcal{B}}}
\def\cD{{\mathcal{D}}}
\def\cG{{\mathcal{G}}}
\def\cF{{\mathcal{F}}}
\def\cM{{\mathcal{M}}}
\def\cN{{\mathcal{N}}}

\def\cR{{\mathcal{R}}}

\def\cT{{\mathcal{T}}}

\def\cI{{\mathcal{I}}}
\def\cJ{{\mathcal{J}}}

\def\cK{{\mathcal{K}}}
\def\cX{{\mathcal{X}}}

\def\cO{{\mathcal{O}}}
\def\cZ{{\mathcal{Z}}}

\def\ga{\gamma}
\def\gae{n_e}
\def\gam{n_m}

\def\sf{{\rm sf}}

\def\tm{{\tilde{m}}}

\def\r2{{\sqrt{2}}}

\def\z{{\zeta}}

\def\th{{\theta}}

\def\cR{{\mathcal{R}}}
\def\w{\omega}

\def\eff{{\rm eff}}
\def\tQ{\tilde{Q}}
\def\tGa{\tilde{\Gamma}}
\def\tq{\tilde{q}}

\begin{document}

\begin{titlepage}

\begin{flushright}
{\bf \today} \\
DAMTP-10-49\\
MAD-TH-10-04\\

\end{flushright}
\begin{centering}

\vspace{.2in}

{ \Large{\bf Notes on Wall Crossing and Instanton in Compactified Gauge Theory with Matter}}

\vspace{.3in}

Heng-Yu Chen${}^{1}$ and Kirill Petunin${}^{2}$\\
\vspace{.2 in}
${}^{1}$Department of Physics, University of Wisconsin, \\
Madison, WI 53706, USA\\
\vspace{.1 in}
and \\
\vspace{.1 in}
${}^{2}$DAMTP, Centre for Mathematical Sciences \\
University of Cambridge, Wilberforce Road \\
Cambridge CB3 0WA, UK \\

\vspace{.2in}

{\bf \large Abstract}
\\
\vspace{.1in}
We study the quantum effects on the Coulomb branch of $\cN=2$ $SU(2)$ supersymmetric Yang-Mills with fundamental matters compactified on $\mathbb{R}^3\times S^1$, and extract the explicit perturbative and leading non-perturbative corrections to the moduli space metric 
predicted from the recent work of Gaiotto, Moore and Neitzke on wall-crossing \cite{GMN1}.
We verify the predicted metric by computing the leading weak coupling instanton contribution to the four fermion correlation using standard field theory techniques, and demonstrate perfect agreement.
We also demonstrate how previously known three dimensional quantities can be recovered in appropriate small radius limit, and provide a simple geometric picture from brane construction.  
\end{centering}

\end{titlepage}

\paragraph{}

\section{Introduction}
\paragraph{}
Understanding the moduli spaces of supersymmetric gauge theories is one of the most physically important and mathematically rich areas in theoretical physics. 
A major triumph in this area is the understanding of exact Coulomb branch metrics for four dimensional $\cN=2$ supersymmetric Yang-Mills, provided by the famous work of Seiberg and Witten \cite{SW1, SW2}. 
The Coulomb branches get perturbative corrections, and also non-perturbative corrections from solitonic objects, in this case Yang-Mills, instantons \footnote{See \cite{InstantonReview} for comprehensive review and reference list.}. However the exact metrics can be extracted elegantly from the periods of the Riemann surfaces for the associated integrable systems.
In this paper, following the recent exciting developments \cite{GMN1,GMN2}, we continue to explore the Coulomb branches for a class of closely related system, namely the compactified $\cN=2$ theories on $\mathbb{R}^3\times S^1$.
In particular, we shall concentrate here on $SU(2)$ gauge group with flavors, related work for pure $SU(2)$ case was done recently in \cite{CDP}.
\paragraph{}
It was known that the Coulomb metrics of the compactified theories are constrained by the eight supercharges to be hyper-K\"ahler, 
and also receive additional non-perturbative soliton corrections, coming from the BPS monopoles/dyons of the uncompactified theories, whose world lines wrap around $S^1$ \cite{SW3D} \footnote{These BPS monopoles/dyons in compactified gauge theories on $\mathbb{R}^3\times S^1$ are thus sometimes referres as ``3d instantons''.}. The Coulomb branch of a compactified theory therefore carries the information about the uncompactified BPS spectrum, in particular, it encodes a subtle but interesting discontinuous jump of the 4d BPS spectrum in the moduli space, known as ``wall-crossing phenomenon'', which we will briefly review in next section.    
In \cite{GMN1}, the authors proposed an exact expression for the Coulumb metrics of the compactified theories, 
and the central object of their construction was a certain ``Darboux coordinate'' $\cX_\ga(\z)$ (\ref{defXgamma}). 
Interestingly, $\cX_{\gamma}(\z)$ is again a solution to a certain Riemann-Hilbert problem with appropriate boundary conditions. It characterizes the moduli space of the famous Hitchin integrable system \cite{GMN2}. 
\paragraph{}
In this note, we are mostly interested in verifying the exact prediction for the Coulomb branch metrics of compactified $\cN=2$ theories with flavors, extracted from the proposed Darboux coordinate $\cX_\gamma(\z)$ \cite{GMN1}. 
The cases with flavors not only serve as another non-trivial confirmation for the proposal in \cite{GMN1}, the knowledge for the explicit Coulomb branch metrics also serve as possible starting point to extend the three dimensional mirror symmetry \cite{3DMirror} to $\mathbb{R}^3\times S^1$. 
We begin with a brief review on the BPS spectrum and wall-crossing phenomenon for $\cN=2$ $SU(2)$ gauge theory with flavors.
Next consider the semi-classical expansion of the proposed wall-crossing integral formula and extract the explicit prediction for the Coulomb branch metric, including both perturbative and non-perturbative corrections. 
The first principles field theory computation for the four fermion correlation function in the monopole background 
is then performed to verify the predicted metric. Along the way, we also explain some subtleties relating to the zero modes and the semi-classical quantization for the monopole in the presence of hypermultiplets. 
We wrap up with explicit demonstration on how the previously known three dimensional metric can be recovered in the $R\to 0$ limit, 
and provide a further geometric understanding from a D-brane picture.

\section{Wall Crossing Formula, Moduli Space and BPS Spectrum with Matter}
\paragraph{}
We consider four dimensional $\cN=2$ supersymmetric gauge theory
with gauge group $G=SU(2)$ and $N_f \le 4$ \footnote{As the $SU(2)$ theories with $N_f>4$ fundamental massless flavors become IR-free.} hypermultiplets in the fundamental representation.
In terms of $\cN=1$ superfield notations, each $\cN=2$ vector multiplet consists of a vector multiplet 
and an adjoint chiral scalar $\Phi$, while
each $\cN=2$ hypermultiplet contains two chiral superfields $Q^i_a$ and $\tilde{Q}_{ia}$, 
($i=1,\dots, N_f$ is the flavor index and $a=1,2$ is the color index.). 
The superpotential preserving $\cN=2$ supersymmetry, including these chiral fields, is given by:
\begin{equation}\label{4DWsuper}
\mathcal{W}_{\rm Matter}=\sum_i \left(\tQ_i \Phi Q^i + m_i \tQ_i Q^i\right)\,,
\end{equation}
where $m_i$ are complex bare masses and the color indices are suppressed. 
This theory has a Coulomb branch $\cB$ where $\phi$, 
the scalar component of $\Phi$ acquires a VEV $\langle\phi\rangle=\frac{a}{2}\sigma_3$, with $\sigma_3$ being Pauli matrix, 
the $SU(2)$ gauge group is broken spontaneously down to $U(1)$.
The massless bosonic fields on the Coulomb branch consist of a $U(1)$ gauge
field and a complex scalar $a$ whose VEV (also denoted as $a$) parametrises
$\cB$ as a complex manifold. It is also convenient to define a
gauge-invariant order parameter $u=\langle {\rm Tr}\, \phi^{2} \rangle$
which provides a globally defined coordinate on $\cB$.
\paragraph{}
The spectrum on the Coulomb branch $\cB$ of the theory contains BPS states
$\gamma=(n_{e},n_{m})$
carrying electric and magnetic charges, $n_e$ and $n_m$, under the
unbroken gauge $U(1)$. Due to additional matter fields, 
the BPS states also transform under the flavor symmetry,
when $m_i=0$, this is $SO(2N_f)$, while $m_i\neq 0$ and are distinct, 
the $SO(2N_f)$ broken down to only $U(1)^{N_f}$,
and the BPS states are labelled by the charges $\{s_i\}$ under $N_f$ $U(1)$s.
We can label a BPS state $Z_{\gamma}(u)$ as\footnote{Here our normalization of the complex bare mass $m_i$ differs that in \cite{SW2} by a factor of $\sqrt{2}$.} 
\begin{equation}\label{DefZgamma}
Z_{\gamma}(u)=\gae a(u)+\gam a_D(u) +\sum_{i}s_i m_i\,,
\end{equation}
which lies on a lattice in the complex plane with periods $a$ and
$a_{D}$. The magnetic period is determined by the
prepotential $\cF(a)$ \cite{SW1} via $a_D=\frac{\partial \cF(a)}{\partial a}$.
The prepotential $\cF(a)$ also determines the low-energy effective
gauge coupling\footnote{In this paper we shall
follow the same normalization convention for
electric charges as in \cite{SW2}, so that the complex gauge coupling $\tau$ is multiplied by a factor of 2 and $a$ is scaled to $a/2$ to compensate.}:
\begin{equation}\label{Deftau}
\tau_{\rm eff}(a)=\frac{8\pi i}{g^2_{\rm eff}(a)}+
\frac{\Theta_{\rm eff}(a)}{\pi}=
\frac{\partial^2
\cF(a)}{\partial a^2}\,.
\end{equation}
In this paper we will mostly be
interested in the weak coupling regime where $|a|\gg|\Lambda|$
($\Lambda$ being the dynamical scale), and we can relate $(a,a_D)$ and
effective coupling $\tau_{\rm eff}$ via
\begin{eqnarray}
a_D\approx \tau_{\rm eff}a\,,
\label{1loop}
\end{eqnarray}
up to exponentially suppressed corrections coming from
four-dimensional Yang-Mills instantons coupling to matters, see \cite{DKM2} for explicit instanton computations.
\paragraph{}
The exact mass formula for BPS states of charge $\gamma$ is
$M_{\gamma}=|Z_{\gamma}|$, with the central charge $Z_{\gamma}$ given
in (\ref{DefZgamma}).  The simplest BPS excitations on the Coulomb
branch $\cB$ are fundamental quarks of electric charge $n_e=+1$,
W-bosons of electric charge $n_e=+2$, and their anti-particles.  The
problem of determining the full BPS spectrum over entire $\cB$ not
only requires $(a(u), a_D(u))$, which were explicitly determined by 
Seiberg and Witten \cite{SW2} in terms of hyperelliptic curves, 
but also the allowed electric-magnetic
charges $(n_e,n_m)$ and the flavor charges $\{ s_i \}$ (if $m_i\neq
0$.).  This amounts to computing the second helicity supertrace
$\Omega(\gamma,u)=-\frac{1}{2}{\rm Tr}_{{\mathcal H} {\rm
BPS},\gamma}(-1)^{2J_3}(2J_3)^2$, which counts the degeneracy of
BPS particles, at each point on the Coulomb branch, where $J_3$ is any
generator of the rotation subgroup of the massive little group. 
For W-bosons this yields
$\Omega(\gamma,u)=-2$, for each fundamental quark of given flavor $\Omega(\gamma,u)=+1$, 
while for monopoles and dyons, which also transform under the flavor group,
their degeneracy factors $\Omega(\gamma,u)$ depend on their explicit representations.
For the cases with massless flavors, the $\Omega(\gamma,u)$ was computed in \cite{DHIZ} from the representations under $SO(2N_f)$, 
for the general massive cases, $\Omega(\gamma,u)$ should depend on the residual $U(1)^{N_f}$ group. 
\paragraph{}
The difficulty in determining $\Omega(\gamma,u)$ in general lies in the existence
of ``walls of marginal stability'' (WMS), they are real codimension
one curves in $\cB$ across which one or more BPS states become
marginally stable. The degeneracies $\Omega(\gamma,u)$ can change
discontinuously as we vary $u$ cross such curves.  In \cite{GMN1},
this phenomenon is described by associating to each BPS state of
charge vector $\gamma$ a ray $l_\gamma$ in a complex spectral plane
with coordinate $\zeta$ \footnote{The auxiliary complex variable
  $\zeta$ is known as the spectral parameter. After adding the point
  at infinity, $\zeta$ parametrises ${\mathbb{CP}}^1$.}:
\begin{equation}\label{Deflgamma}
l_\gamma :=\left\{\zeta: \frac{Z_\gamma(u)}{\zeta}\in \bR_{-}\right\}\,.
\end{equation}
These BPS rays rotate in the $\zeta$-plane as we move in $\cB$.  On
the wall of marginal stability, a set of BPS rays $\{Z_\gamma(u)\}$
become aligned, and their charges can be parametrised as
$\{N_1\gamma_1+N_2\gamma_2\},~N_1,N_2 > 0$, for some primitive vectors
$\gamma_{1,2}$ with $Z_{\gamma_1}/Z_{\gamma_2} \in \bR_+$ \cite{GMN1}.
The condition that BPS rays become aligned is equivalent to satisfying
the energy conservation, the conservations of electric and magnetic charges
$(n_e,n_m)$, and flavor charges $\{s_i\}$.
\paragraph{}
It is well-known in the literature \cite{Henn95, BF1, BF2} that the ${\cal N}=2$ theory with gauge group $SU(2)$ and $N_f$ fundamental matter 
has non-trivial curves of marginal stability and exhibit the wall-crossing phenomenon. 
In the purely massless cases $m_i=0$ for $N_f=1,2,3$ \footnote{Note that for massless case, $SU(2)$ with $N_f=4$ is conformal, the BPS spectrum can be purely determined in the semi-classical regime as in \cite{GaHa}.}, 
the curve of marginal stability is given by the locus ${\rm Im}(a_D(u)/a(u))=0$ in $\cB$, which can be solved numerically from explicit profile of $(a(u),a_D(u))$. 
Such curve divides up the Coulomb branch $\cB$ into weakly and strongly coupled regimes, and goes through singular points where BPS particles can become massless, the BPS spectra are different inside and outside the curve. 
In the weak-coupling region
where $|u|\gg|\Lambda|^2$, the BPS spectrum for $N_f=1,2,3$ consists of the fundamental (anti-) quarks of charges $\pm (1,0)$, 
W-bosons of charges $\pm(2,0)$ and tower of dyons $\pm(n,1)\,, n\in {\mathbb Z}$, 
in addition for $N_f=3$ there are also dyons of charges $\pm(2n+1,2)\,, n\in{\mathbb Z}$ 
\footnote{For completeness here we also mention $N_f=4$ case \cite{GaHa, DHIZ}, its semi-classical BPS spectrum, as classified by representation under $SO(8)$ flavor symmetry group are:  $(2n,2m)$ as singlet ${\bf 1}$; $(2n+1, 2m)$ as vector ${\bf 8_v}$; $(2n,2m+1)$ as spinor ${\bf 8_s}$; and $(2n+1,2m+1)$ as conjugate ${\bf 8_c}$. Here integers $n,m$ are co-prime.}. 
All of them transform under $SO(2N_f)$ flavor groups and their representations are summarized in the section 3 of \cite{DHIZ}.
As we enter the strongly coupled region near the origin in $\cB$ and cross the wall of marginal stability, most states of the
semiclassical BPS spectrum decay into only finite number of stable BPS states, as explicitly determined in \cite{BF1}.
\paragraph{}
When we further include bare masses $m_i \neq 0$ into the theories,
the set of walls of marginal stability becomes very complicated \cite{BF2}, as they contribute further parameters into the problem
\footnote{It also has to be said that a systematic determination of BPS spectrum for general four dimensional $\cN=2$ $SU(N), N>2$ theories is currently lacking.}.
Until recently,
a systematic determination of the BPS spectrum for generic $\cN=2$
SUSY gauge theories and all regions of the moduli space remained elusive.
Recently significant progress has been made in the work of Gaiotto, Moore and Neitzke
towards a systematic determination of the BPS spectra for four dimensional $\cN=2$ supersymmetric gauge theories \cite{GMN1,GMN2}. 
In particular, following the work by Kontsevich and Soibelman \cite{KS1,KS2}, 
they proposed an explicit wall-crossing formula, which in principle encodes the discontinuous BPS spectrum across the wall of marginal stability.
Here we are mostly interested in the consequences of their conjecture for
the $\cN=2$ theory with fundamental flavors. 
To apply the basic idea in \cite{GMN1} to our case, we first Euclideanize and compactify it on $\bR^3\times S^1$,
where $S^1 : x^4 \sim x^4+2\pi R$ has radius $R$. On length-scales much larger
than $R$, the low energy effective action on the Coulomb branch becomes three-dimensional. In
addition to the four dimensional complex scalar $a$,
two additional real periodic scalar fields now appear.
The first one is the electric Wilson line, which comes from the
component $A_{4}$ of the $U(1)$ gauge along $S^1$; the other one comes from dualizing the three dimensional $U(1)$ Abelian gauge
field in favor of another real scalar,
$\theta_{m}\in [0,2\pi]$, known as the ``magnetic Wilson line''.
We can denote them as:  
\begin{equation}\label{Defthetae}
\theta_e=\oint_{S^1} A_4 dx^4\,,~~~\theta_m=\oint_{S^1}(A_{D,4}) dx^4\,,
\end{equation}
with $(\theta_e,\theta_m) \sim (\theta_e+2\pi, \theta_m+2\pi)$, they
describe a two-torus.  Turning to the matter sector, in four dimensions
the bare mass $m_i$ of a hypermultiplet is a complex parameter with
two real components, but when we compactify the theory on $S^1$, a
third real component $\tm_i$ called ``real mass'' appears.  To see
this, we can view the masses $m_i$ and $\tm_i$ as the expectation
values of background vector multiplets which weakly gauge flavor
symmetries, this also makes clear that they should transform in the
adjoint of $SO(2N_f)$ \cite{APS}.  The supersymmetry condition
requires that the masses can only be gauged in maximal torus of $SO(2N_f)$,
the real mass $\tm_i$ then appears as the Wilson line VEV under the
$S^1$ compactification. In the brane picture discussed in
section~\ref{sec:3d-and-branes}, this real mass has a simple
interpretation as the separation of the gauge and flavor branes in the
dual compactified dimension.
\paragraph{}
The Coulomb branch of the compactified theory is a four real
dimensional manifold ${\cal M}$, paramet\-rised by the scalars
$\{a,\bar{a},\theta_{e},\theta_{m}\}$.  The low-energy effective field
theory on the Coulomb branch is then given by a three dimensional sigma model with
target $\cM$, the eight supercharges further constrain $\cM$ to be
hyper-K\"ahler \cite{HKLR}.  It is well-known that a hyper-K\"ahler
manifold admits a ${\mathbb{CP}}^1$ worth of complex structures,
coming from $SU(2)$ triplet of complex structures $J_{1,2,3}$.  We can
use the complex variable $\zeta$ to parametrise such ${\mathbb{CP}}^1$
(same as the $\zeta$ appearing in (\ref{Deflgamma})).  The so-called
twistor space $\cT$ of a hyper-K\"ahler manifold is constructed
to incorporate all possible complex structures, and it is given
topologically by $\cT \sim \cM \times {\mathbb{CP}}^1$ \cite{HKLR}. 
In particular, choosing the complex
coordinates holomorphic with respect to $J_3$, we can now organise the
general K\"ahler form as
\begin{equation}\label{Defwzeta}
\w(\zeta)=-\frac{i}{2\z}\w_+ +\w_3-\frac{i}{2}\z \w_-\,,
\end{equation}
where $\w_{1,2,3}$ are the K\"ahler forms associated with $J_{1,2,3}$
and we have defined $\w_{\pm}=\w_1\pm i\w_2$, the metric $g$ is then
extracted from the $\zeta$ independent component of $\omega(\zeta)$.
The twistor space $\cT$ plays an important role in the construction of
the metric on $\cM$ in \cite{GMN1}, however for the purpose of this
paper, it is sufficient to regard it as an auxiliary space.
\paragraph{}
We can deduce the leading order low-energy effective action from direct 
dimensional reduction of the four-dimensional low-energy
theory. To describe the action we define the complex combination
$z=\th_m-\tau_{\rm eff}\th_e$ which parametrises a torus
with complex structure $\tau_{\rm eff}(a)$.
In this limit the moduli space $\cal{M}$ corresponds to a fibration of
this torus over the Coulomb branch $\cB$ of the four-dimensional
theory. The real bosonic part of the
resulting
action is given in terms of scalar fields $\{a,\ba,\th_e,\th_m\}$ as
\begin{eqnarray}
\label{bosonic 3D action} \label{DefL3sf}
S_{\rm B}&=&\frac{1}{4}\int d^3 x \left(\frac{8\pi R}{g_{\rm eff}^2}
\partial_\mu a\,\partial^\mu\bar{a} +  \frac{g^2_{\rm eff}}{32\pi^3
R}\partial_\mu z\,\partial^\mu \bar
z\right)\,. \end{eqnarray}
In addition, surface terms give rise to pure imaginary
terms in the action depending on the total electric and magnetic
charges,
\begin{eqnarray}
S_{\rm Im} & = & i\left(n_{e}\,\,+\,\, \frac{\Theta_{\rm eff}}
{\pi}\,n_{m}\right ) \,\theta_{e}\,\,+\,\,in_{m}\theta_{m}\,.
\label{Im}
\end{eqnarray}
The term proportional to $\Theta_{\rm eff}$ arises from dimensional
reduction of the $F\wedge F$ term in the low-energy action of the
four-dimensional theory after replacing $A_{4}$ by $\theta_{e}/2\pi
R$. As explained in \cite{SW2}, the $(n_e+n_m\frac{\Theta_{\rm
    eff}}{\pi})$ term is the eigenvalue of the electric charge
operator, we should regard it as effective electric charge.  The
corresponding fermionic terms in the action take the form
\begin{eqnarray}
\label{fermionic 3D action}
S_{\rm F}&=&\frac{4\pi R}{g^2_{\rm eff}}\int d^3
x\left(i\bar\psi\bar\sigma^\mu\partial_\mu\psi+i\bar
\lambda\bar\sigma^\mu\partial_\mu \lambda\right)\,,
\end{eqnarray}
where $\lambda$ and $\psi$ are the dimensional reduction along the
$x_{4}$ direction of the four-dimensional Weyl fermions in the $U(1)$
vector multiplet whose lowest component is the scalar $a$.
The leading order effective Lagrangian (\ref{DefL3sf}) allows us to 
extract the leading $R\to \infty$ behavior of the hyper-K\"ahler metric
on $\cM$,
\begin{equation}\label{Defgsf}
g^{\rm sf}=R({\rm Im\, \tau_{\rm eff}})|da|^2+\frac{1}{4\pi^2 R}({\rm
Im}\, \tau_{\rm eff})^{-1}|dz|^2\,.
\end{equation}
As $\{\th_e,\th_m\}$ span a flat two torus, (\ref{Defgsf}) was
referred to as the ``semi-flat'' metric in \cite{GMN1}.  The metric
(\ref{Defgsf}) also makes apparent that $g^{\rm sf}$ is K\"ahler with
respect to the complex structure where $\{a,z\}$ are holomorphic
coordinates. Going to finite radius $R$, this semi-flat metric gets
quantum corrected by perturbative one-loop corrections 
and a series of instantons coming from
the four-dimensional BPS states whose worldlines now wrap around
$S^1$. We shall discuss them in turns.
\paragraph{}
The authors of \cite{GMN1}, in addition to providing a method for
determining the four dimensional BPS spectrum across the wall of
marginal stability, also predict the smooth fully quantum-corrected
metric $g$ for the Coulomb branch of the compactified theory on
$\mathbb{R}^3\times S^1$.  The metric is effectively determined once the
one-parameter family of K\"ahler forms $\w(\zeta)$ introduced above is
known \footnote{Here our convention relating the K\"ahler form
  $\omega$ and metric $g$ is such that $\omega = i\,\partial^2
  K/(\partial z^{a} \partial z^{\bar b})d z^{a}\wedge d z^{\bar b}$
  and $g=2\,\partial^2 K/(\partial z^{a} \partial z^{\bar b})d z^{a}
  dz^{\bar b}=2g_{a\bar b}d z^{a} d z^{\bar b}$, where $K$ is the
  K\"ahler potential.}.  For any complex symplectic manifold, which is a
hyper-K\"ahler manifold, one can always find Darboux coordinates
locally in which the symplectic form becomes canonical.  In the
present case we introduce complex coordinates $\mathcal{X}_{e}(\zeta)$
and $\mathcal{X}_{m}(\zeta)$, in terms of which
\begin{equation}\label{Defwzeta2}
\omega(\zeta)=-\frac{1}{4\pi^2 R}\frac{d\mathcal{X}_{e}}{\mathcal{X}_{e}}
\wedge
\frac{d\mathcal{X}_{m}}{\mathcal{X}_{m}}\,.
\end{equation}
More generally, we also introduce a corresponding Darboux coordinate
$\cX_\gamma(\z)$ associated with any vector $\gamma$ in the
charge lattice determined by the relation
$\mathcal{X}_{\gamma_1+\gamma_2}=
\mathcal{X}_{\gamma_1}\mathcal{X}_{\gamma_2}$ where
$\mathcal{X}_{\gamma}= \mathcal{X}_{e}$ for $\gamma=(1,0)$ and
$\mathcal{X}_{\gamma}= \mathcal{X}_{m}$ for $\gamma=(0,1)$.
In the large-$R$ limit, the semi-flat metric (\ref{Defgsf})
corresponds to the choice,
\begin{equation}\label{defXsf}
\cX_{\gamma}^{\sf}(\z)=\exp\left(\pi R \frac{n_e a+n_m a_D}{\z}+i(n_e\theta_e + n_m \theta_m )+\pi R\z \overline{(n_e a+n_m a_D)}
\right)\,.
\end{equation}
It turns out that this asymptotic behavior, along with the requirement
of continuity of $\mathcal{X}_{\gamma}(\z)$ across walls of marginal
stability (as governed by the Kontsevich-Soilbelman algebra) is enough
to determine $\mathcal{X}_{\gamma}(\z)$, and hence the metric on
$\cM$, at any point on the complex $\zeta$-plane. In the case of a
theory with flavors, it is necessary to include the mass contributions
$m_i$ and $\tm_i$. The relevant expression for the Darboux coordinate
$\cX_\gamma(a,\th,\z)$ is given by the following integral equation \footnote{Wall Crossing phenomenon for $SU(2)$ with fundmamental flavors has also been considered in \cite{DGS}.}
\cite{GMN1}:
\begin{equation}\label{defXgamma}
\cX_{\ga}(\z)=
\cX^{\sf}_\ga(\z)\exp\left[-\frac{1}{4\pi i}
\sum_{\ga'\in\Gamma}\Omega(\ga';u)\langle\ga,\ga'\rangle\int_{l_{\ga'}}\frac{d\z'}{\z'}\frac{\z'+\z}{\z'-\z}\log\left(1-\sigma(\ga')\cX_{\ga'}(\z')\mu_\ga(\z')\right)\right]\,,
\end{equation}
with \footnote{Here we are following the central charge convention used in \cite{GMN1}, 
such that $\cX_{\ga}^{\sf}(\z)\mu_\ga(\z)=\exp(\pi R\z^{-1} Z_\ga+\pi R \z \bar{Z}_\ga+\dots)$, where $Z_\ga$ is the central charge including complex mass $m_i$ contribution as defined in (\ref{DefZgamma}).}
\begin{equation}\label{Defmu}
\mu_\gamma(\z)=\exp\left[\sum_{i}^{N_f}s_i\left(\frac{\pi R}{\z}m_i+i {\psi_i} + \pi R{\z}{\bar{m}_i}\right)\right]\,,~~~\psi_i=2\pi R \tm_i \,.
\end{equation}
In above we have also introduced the following quantities:
$\langle\ga,\ga'\rangle$ is the symplectic product between two charge
vectors, $\gamma=(\gae,\gam)$ and $\gamma'=(\gae',\gam')$,
which we can take to be
\begin{equation}
\langle\ga,\ga'\rangle =
\langle(\gae,\gam),(\gae',\gam') \rangle =-\gae\gam'+\gae'\gam\,,
\end{equation}
and the ``quadratic refinement'' $\sigma(\ga')$ is given by
$\sigma(\ga')=(-1)^{\gae'\gam'}$.  The summation in (\ref{defXgamma})
is over the set of charges $\Gamma$ in the theory and the integration
contour $l_{\ga'}$ associated with $\gamma'$ is the BPS ray as defined
in (\ref{Deflgamma}), this ensures the convergence of the integral.

\section{Semiclassical Limit of the Wall-Crossing Formula with Matters}
\paragraph{}
To solve (\ref{defXgamma}), we can take the logarithm and see that the
right hand side contains a source term corresponding to the semi-flat
expression and an integral convolution.  To extract the explicit
prediction for the quantum corrected moduli space metric from
(\ref{defXgamma}), here we generalize the iterative weak coupling
expansion, first done in \cite{CDP} for the pure $SU(2)$ case, to the
current case with massive flavors.  We therefore restrict our
attention to the semiclassical region of the moduli space, such that
$|a|\gg\Lambda$ and $g^{2}_{\rm eff}\ll1$, while keeping fixed
the dimensionless quantity $R|a|$. As we explain below, the
quantity $\exp(-2\pi R |Z_{\gamma}|)$ is then exponentially suppressed
for all states with non-zero magnetic charge.  For the weak coupling
spectrum described above, this is the case for all BPS states except
the purely electrically charged fundamental quarks and massive gauge
bosons, they essentially contribute the perturbative one-loop
corrections to the moduli space metric.
\paragraph{}
We begin by decomposing the Darboux coordinate $\cX_\ga(\z)$ as
$\cX_\ga(\z)=\left[\cX_e(\z)\right]^{\gae}\left[\cX_m(\z)\right]^{\gam},~\gamma=(\gae,\gam)$.
The integral equation (\ref{defXgamma})
for the electric and the magnetic Darboux coordinates is then given
as
\begin{eqnarray}
\cX_e(\z)&=&\cX_e^{\sf}(\z)\exp\left[-\frac{1}{4\pi
i}\sum_{\ga'\in\Gamma}c_e(\ga')\,\cI_{\ga'}(\z)\right]\,,
~~~c_e(\ga')=\Omega(\ga';u)\langle(1,0),\ga'\rangle\,,\label{DefXe}\\
\cX_m(\z)&=&\cX_m^{\sf}(\z)\exp\left[-\frac{1}{4\pi
i}\sum_{\ga'\in\Gamma}c_m(\ga')\,\cI_{\ga'}(\z)\right]\,,
~~~c_m(\ga')=\Omega(\ga';u) \langle(0,1),\ga'\rangle\,,\label{DefXm}
\end{eqnarray}
where $\cX_{e}^{\sf}(\z)$ and $\cX_{m}^{\sf}(\z)$ are given by (\ref{defXsf})
with $(n_e,n_m)$ equal to $(1,0)$ and $(0,1)$ respectively, and
$\cI_{\ga'}(\z)$ is defined to be
\begin{equation}\label{DefIzeta}
\cI_{\gamma'}(\z)=\int_{l_{\ga'}}\frac{d\z'}{\z'}\frac{\z'+\z}{\z'-\z}
\log(1-\sigma(\ga')\mu_{\ga'}(\z')\cX_{\ga'}(\z'))\,.
\end{equation}
Taking the weak coupling limit, which sets $a_D\approx \tau_{\rm eff} a$ up to one loop order, we find
\begin{equation}\label{appXsf}
\log\cX_e^{\sf}(\z)=\pi R a\z^{-1}+i\theta_e+\pi R \ba \z\,,
~~~\log\cX_m^{\sf}(\z)=\pi R a\tau_{\rm eff}(a)\z^{-1}+i\theta_m+\pi R
\overline{a\tau_\eff(a)} \z.
\end{equation}
We can see that in this limit $\log|\cX^{\sf}_m|\gg \log|\cX_e^{\sf}|$, this
has interesting consequences for deriving an iterative solution to
$\cX_\ga(\z)$.
We can explicitly expand $\log\cX_e(\z)$ and $\log \cX_m(\z)$ for the weak
coupling spectrum of $SU(2)$ theories with $N_f$ flavors discussed earlier:
\begin{eqnarray}
&&\log \cX_e(\z)=\log\cX_e^{\sf}(\z)-\frac{1}{4\pi i}\sum_{\ga' \in \tGa}c_e(\ga')\cI_{\ga'}(\z)\,,\label{logXe}\\
&&\log \cX_m(\z)=\log\cX_m^{\sf}(\z)-\frac{1}{4\pi i}\sum_{\ga'\in \{W^{\pm}, q_i^{\pm}, \tq_i^{\pm}\}} c_m(\ga')\cI_{\ga'}(\z)
-\frac{1}{4\pi i}\sum_{\ga'\in \tGa}c_m(\ga')\cI_{\ga'}(\z)\,.
\label{logXm}
\end{eqnarray}
Here in our expansion for $\cX_m(\z)$ we have singled out the purely electrically charged BPS states $\{W^{\pm}, q_i^{\pm}, \tq_i^{\pm}\}$, 
the summation over $N_f$ flavors of fundamental quarks is also implied. 
The set $\tGa$ denotes the remaining weakly coupled BPS spectrum after omitting the purely electrically charged ones, in other words those ones that are magnetically charged.
They act as non-perturbative instanton corrections to the Coulomb branch metric on $\mathbb{R}^3\times S^1$. 
At weak coupling, the central charge takes the form $Z_\ga(a)=
a(\gae+\gam \tau_{\rm eff}(a)) +\sum_{i=1}^{N_f}
s_i{m_i}$,  
where $\tau_{\rm eff}=\frac{8\pi i}{g_{\rm eff}}+\frac{\Theta_{\rm eff}}{\pi }$ 
with $g_\eff$ and $\Theta_\eff$ now denote the effective coupling constant and
the effective vacuum angle. 
\paragraph{}
To solve for $\log \cX_{\ga}$ iteratively, at the leading order, 
we substitute the semi-flat coordinates (\ref{appXsf})
into the right hand side of (\ref{logXe}, \ref{logXm}) and ignore the
components which vanish as $g_\eff\rightarrow 0$.
In such a limit, BPS contributions from $\tGa$ to $(\cX_e, \cX_m)$ duly vanish (i.e. the last term in (\ref{logXe}) and (\ref{logXm})), 
$\cX_m$ does however receive order one contributions from purely electrically charged $\{W^{\pm},q_i^{\pm},\tq_i^{\pm}\}$, which we shall proceed to compute momentarily.
We shall denote the resultant coordinates at this order as
$(\cX^{(0)}_e,\cX^{(0)}_m)$. Thus we have
\begin{eqnarray}
\log\cX_e^{(0)}(\z)&=&
\log\cX_e^{\sf}(\z)\,,~~~\log\cX_m^{(0)}(\z)=\log\cX_m^{\sf}(\z)+\log\cD(\z)\,,\label{DefXe0Xm0}\\
\log\cD(\z)&=& \log\cD_W(\z)+\log \cD_{q}(\z)+\log \cD_{\tq}(\z)\,.
\label{DefDz}
\end{eqnarray}
Here we have split the electric contributions $\cD(\z)$ to $\cX_m(\z)$ into three pieces:
$\cD_W(\z)$, coming from W-bosons, and $\cD_{q}(\z)$ and $\cD_{\tq}(\z)$, coming from the quarks in fundamental hypermultiplet.
They are defined in turns as:
\begin{equation}
\log \cD_{W}(\z)= \frac{1}{\pi i} \left( \int_{l_{W^+}} \frac{d\z'}{\z'}\frac{\z'+\z}{\z'-\z}\log\left[1-\cX_{W^+}^{\sf}(\z')\right]
-\int_{l_{W^-}} \frac{d\z'}{\z'}\frac{\z'+\z}{\z'-\z}\log\left[1-\cX_{W^-}^{\sf}(\z')\right] \right)\,,\label{DW}
\end{equation}
\begin{equation}
\begin{aligned}
\log \cD_{q}(\z)= -\frac{1}{4\pi i}\sum_{i=1}^{N_f}\left(\int_{l_{q_i^+}} \frac{d\z'}{ \z'}\frac{\z'+\z}{\z'-\z}\log\left[1-\mu_{q^+_i}(\z')\cX_{q^+_i}^{\sf}(\z')\right] \right. \\ \left.
-\int_{l_{q_i^-}} \frac{d\z'}{\z'}\frac{\z'+\z}{\z'-\z}\log\left[1-\mu_{q^-_i}(\z')\cX_{q^-_i}^{\sf}(\z')\right]\right) \,,\label{Dq}
\end{aligned}
\end{equation}
\begin{equation}
\begin{aligned}
\log \cD_{\tq}(\z)= -\frac{1}{4\pi i}\sum_{i=1}^{N_f}\left(\int_{l_{\tq_i^+}} \frac{d\z'}{ \z'}\frac{\z'+\z}{\z'-\z}\log\left[1-\mu_{\tq_i^+}(\z')\cX_{\tq_i^+}^{\sf}(\z')\right] \right. \\ \left.
-\int_{l_{\tq_i^-}} \frac{d\z'}{\z'}\frac{\z'+\z}{\z'-\z}\log\left[1-\mu_{\tq^-_i}(\z')\cX_{\tq^-_i}^{\sf}(\z')\right]\right) \,.
\label{Dtq}
\end{aligned}
\end{equation}
In evaluating (\ref{DW})-(\ref{Dtq}), we have used $\Omega(W^{\pm},u)=-2$ and $\Omega(q_i^{\pm},u)=\Omega(\tq_i^{\pm},u)=1$, and their charges, $\pm (2,0)$ for $W^{\pm}$ and $\pm(1,0)$ for $q^{\pm}_i$ and $\tq^{\pm}_i$.  
The mass parameters are given by $\mu_{q^{\pm}_i}(\z)=\exp[\pm({m_i}\z^{-1}+i\psi_i+{\bar{m_i}}\z)]$
and $\mu_{\tq^{\pm}_i}(\z)=\exp[\mp({m_i}\z^{-1}+i\psi_i+\bar{m_i}\z)]$ respectively, with $\psi_i=2\pi R\tm_i$.
In other words for fundamental hypermultiplets $q_i^{\pm}$ and $\tq_i^{\pm}$, the flavor charges are $s_j=\pm \delta_{ij}$ and $s_j= \mp \delta_{ij}$ respectively, while for the W-bosons $W^{\pm}$, which are in the vector multiplet, $s_i=0$.
\paragraph{}
We can now further expand $(\cX_e(\z),\cX_m(\z))$ to extract the non-perturbative corrections:
\begin{equation}\label{XeXm0exp}
\log\cX_{e}(\z)\,\,\,=\,\,\,\log\cX_{e}^{(0)}(\z)\,\,+\,\,\delta
\log\cX_e(\z)\,,
\quad{}
\log\cX_{m}(\z)\,\,\,=\,\,\,\log\cX_{m}^{(0)}(\z)\,\,+\,\,\delta
\log\cX_m(\z)\,.
\end{equation}
We can compute $(\delta \cX_{e}(\z),\delta \cX_m(\z))$ by substituting $(\cX^{(0)}_e(\z),\cX^{(0)}_m(\z))$ into (\ref{logXe}) and (\ref{logXm}):
\begin{eqnarray}
\delta\log\cX_e(\z)&=&-\frac{1}{4\pi i}\sum_{\ga'\in \tGa}
c_e(\ga')\cI_{\ga'}^{(0)}(\z)\,,\label{delogXe0}\\
\delta \log\cX_m(\z)&=&-\frac{1}{4\pi i}\sum_{\ga'\in \tGa} c_m(\ga')\cI_{\ga'}^{(0)}(\z)\,,
\label{delogXm0}
\end{eqnarray}
where we have defined the short-hand notation for the integral:
\begin{equation}
\cI_{\gamma'}^{(0)}(\z)
=\int_{l_{\ga'}}\frac{d\z'}{\z'}\frac{\z'+\z}{\z'-\z}
\log\left[1-\sigma(\ga')\mu_{\ga'}(\z')\left(\cX_e^{(0)}(\z')\right)^{\gae'}\left(\cX_m^{(0)}(\z')\right)^{\gam'}\right]\,.\label{DefI0}
\end{equation}
The BPS ray $l_{\gamma'}$ which plays the role of integration contour is defined in (\ref{Deflgamma}) with $Z_{\gamma'}$ as given in (\ref{DefZgamma}).
Substituting the corrected Darboux coordinates (\ref{XeXm0exp}) 
into the definition of the symplectic form $\w(\z)$ (\ref{Defwzeta2}), we can find the corresponding corrections to the metric on $\cM$,
\begin{eqnarray}\label{omegaexpansion}
\w(\z)
&=& \w^{\sf}(\z)+\w^{\rm P}(\z)+\w^{\rm NP}(\z)+{\mathcal{O}}(\delta^2)\,.
\end{eqnarray}
The various terms in (\ref{omegaexpansion}) are given explicitly by
\begin{eqnarray}
\w^{\sf}(\z)&=&-\frac{1}{4\pi^2 R}d\log\cX_e^{\sf}(\z)\wedge d\log
\cX_m^{\sf}(\z)\,,\label{wsf}\\
\w^{\rm P}(\z)&=&-\frac{1}{4\pi^2 R}d\log\cX_e^{\sf}(\z)\wedge d\log
\cD(\z)\,,\label{wW}\\
\w^{\rm NP}(\z)&=&-\frac{1}{4\pi^2 R}\left(d\delta\log\cX_e(\z)\wedge
d\log\cX_m^{(0)}(\z)+d\log\cX_e^{(0)}(\z)\wedge
d\delta\log\cX_m(\z)\right)\,.\label{winst}
\end{eqnarray}
The term $\w^{\rm P}(\z)$ corresponds to the one-loop perturbative
corrections to the metric due to the $W^{\pm}$ bosons and the quarks $q_i^{\pm}$ and $\tq_i^{\pm}$, 
and they can be readily evaluated using modified Bessel functions of second kind $K_{\nu}(x)$:
\begin{eqnarray}
\w^{\rm P}(\z)&=&-\frac{i}{4\pi^2 R}d\log\cX_e^{\sf}(\z)\wedge \left[2\pi 
A^{\rm P}(a,\ba)+\pi  V^{\rm P}(a,\ba)(\z^{-1} da-\z d\ba)\right]\label{expwW}\,,\\
A^{\rm P}(a,\ba)&=&\frac{R}{2\pi}\sum_{k>0} \sum_{\ga\in\{W^{\pm},q_i^{\pm},\tq^{\pm}_i\}} n_{e} c_m(\gamma)|Z_\gamma| 
e^{ik(\th_\ga+s_i\psi_i)} K_1(2\pi R |k Z_\ga|)
\left(\frac{da}{Z_\ga}-\frac{d\ba}{\bar{Z}_\gamma}\right)\,,\label{AW}\\
V^{\rm P}(a,\ba)&=&-\frac{R}{\pi}\sum_{k> 0} \sum_{\ga\in \{W^{\pm},q_i^{\pm},\tq^{\pm}\}} n_e c_m(\gamma) e^{ik(\th_\ga+s_i\psi_i)}K_0(2\pi
R |k Z_\ga|)\,,\label{VW}
\end{eqnarray}
Using $n_e=\pm 2$ for $W^{\pm}$ and $n_e=\pm 1$ for $q_i^{\pm}$ and $\tq_i^{\pm}$,
we can obtain that $c_m(W^{\pm})=\mp 4$ and $c_m(q_i^{\pm})=c_m(\tq_i^{\pm})=\pm 1$. 
Notice that here we are taking weak coupling limit while
keeping $R |a|$ fixed and arbitrary, as first non-trivial check 
one can consider taking three dimensional limit $R|a|\to 0$.  
To do so, first notice that $K_{\nu}(x)$ diverges when $x\to 0$, 
we should Poisson resum the series of Bessel functions over $k$, which is equivalent to summing over all the Kaluza-Klein momentum modes.
When combining with the leading semi-flat piece,
we can extract the shift of the coupling constant from the moduli space metric:
\begin{eqnarray}\label{DefMn}
&&\frac{8\pi R}{g^2_{\rm eff}}\rightarrow\frac{8\pi R}{g_{\rm eff}^2}-
\frac{1}{8\pi}\sum_{n\in \bbZ}\left[\frac{8}{
|M_W(n)|}-\sum_{i=1}^{N_f}\left(\frac{1}{|M_{q_i}(n)|}+\frac{1}{|M_{\tq_i}(n)|}\right)\right]\,,\\
&&|M_W(n)|=\sqrt{|2a|^2+\left(\frac{\theta_e}{\pi R}+\frac{n}{R}\right)^2}\,,\\
&&|M_{q_i}(n)|=\sqrt{\left|a+{m_i}\right|^2+\left(\frac{\theta_e+\psi_i}{2\pi R}+\frac{n}{R}\right)^2}\,,~~~
|M_{\tq_i}(n)|=\sqrt{\left|a-{m_i}\right|^2+\left(\frac{\theta_e-\psi_i}{2\pi R}+\frac{n}{R}\right)^2}\,,\nonumber\\
\end{eqnarray}
where in $R\to 0$ limit, all the KK-momentum modes decouple for $n\neq 0$. 
For single flavor $N_f=1$ case, this precisely coincides with the Coulomb branch metric predicted by \cite{SW3D}, which is the three-parameters family $({\rm Re}(m), {\rm Im}(m), \tm)$ deformation of the double cover of Atiyah-Hitchin manifold discovered by Dancer \cite{Dancer}.
For generic $N_f$, the shift also matches with the first principle one-loop computations performed in \cite{DTV}, after taking into account of the normalization of gauge coupling and electric charges.
\paragraph{}
For the non-perturbative contributions $\w^{\rm NP}(\z)$, 
as explained in detail in \cite{CDP}, as the expression is dominated by the exponential factors at weak coupling,
they can be readily evaluated using a saddle point approximation:
\begin{eqnarray}
\w^{\rm
\rm NP}(\z)&=&\sum_{\gamma\in \tGa}\Omega(\ga,u)\w_{\ga}(\z)\,,\label{wNP}\\
\w_{\ga}(\z)&=&-\frac{1}{4\pi^2
R}\frac{d\cX_{\ga}^{(0)}(\z)}{\cX_{\ga}^{(0)}(\z)}\wedge\left(\sum_{k=1}^{\infty}\frac{1}{4\pi i
}\int_{l_{\ga}}\frac{d\z'}{\z'}\frac{\z'+\z}{\z'-\z}
\left[\sigma(\gamma)\mu_{\ga}(\z')\cX^{(0)}_{\ga}(\z')\right]^k \frac{d\cX_{\ga}^{(0)}(\z')}{\cX_{\ga}^{(0)}(\z')}\right)
\nonumber\\
&\approx & -\frac{1}{4\pi^2
R}\frac{d\cX_{\ga}^{\sf}(\z)}{\cX_{\ga}^{\sf}(\z)}\wedge\left(\sum_{k=1}^{\infty}\frac{1}{4\pi i
}\int_{l_{\ga}}\frac{d\z'}{\z'}\frac{\z'+\z}{\z'-\z}\left[\sigma(\gamma)\mu_{\ga}(\z')\cX^{(0)}_{\ga}(\z')\right]^k
\frac{d\cX_{\ga}^{\sf}(\z')}{\cX_{\ga}^{\rm
sf}(\z')}\right)\,.
\label{Defwemk}
\end{eqnarray}
Here we have used the fact that along each integration contour
$l_{\gamma} : Z_{\gamma}/\z' \in {\mathbb{R}_{-}}$, the zeroth order Darboux
coordinate
$\mu_{\ga}(\z')\cX^{(0)}_{\ga}(\z')=\mu_{\ga}(\z')\cX^{\sf}_{\ga}(\z')\cD(\z')$ is proportional to
exponential factor
$\exp\left[-\pi R|Z_{\ga}|(|\z'|+1/|\z'|)\right]$, which ensures the
convergence of the integral.\footnote{Here we have also further approximated $d\cX^{(0)}(\z)/\cX^{(0)}(\z)$ by
$d\cX_{\ga}^{\sf}(\z)/\cX_{\ga}^{\sf}(\z)$,
as the contribution proportional to $d\cD(\z)/\cD(\z)$ is of higher order in
$g^2_{\rm eff}$ in our saddle point analysis.}
As shown in \cite{CDP}, at weak coupling the saddle point is at $\z'=-Z_{\ga}/|Z_{\ga}|\approx -i[n_m]$, 
where $[n_m]$ denotes the sign of the magnetic charge for $\ga$.  
Upon substitution and performing the Gaussian fluctuation integral,
the leading expression for $\w_{(\ga',k)}(\z)$ is given by
\begin{eqnarray}
\w_{\ga}(\z)&=&\sum_{k=1}^{\infty}\w_{(\ga,k)}(\z)\,,\label{wdyon1}\\
\w_{(\ga,k)}(\z)&=&
\cJ_{(\ga,k)}\frac{d\cX_{\ga}^{\sf}(\z)}{\cX_{\ga}^{\sf}(\z)}\wedge
\left[{|Z_{\ga}|}\left(\frac{dZ_{\ga}}{Z_{\ga}}-\frac{d\bar{Z}_{\ga}}{\bar{Z}_{\ga}}\right)
-\left(\frac{dZ_{\ga}}{\z}-\z d\bar{Z}_{\ga}\right)\right]\, ,\\
\cJ_{(\ga,k)}&=& -\frac{1}{16\pi^2 i }
\frac{\cD(-i [n_m])^{k\gam}}{\sqrt{k R |Z_{\ga}|}} \left[\sigma(\gamma)\right]^{k}
\exp\left[k\left(-2\pi R|Z_{\ga}|+i\left(\th_{\ga}+\sum_{i=1}^{N_f} s_i \psi_i\right)\right)\right]\,.\label{sadcJgmk}
\end{eqnarray}
The one loop determinant $ \cD(-i[n_m])=\cD_W(-i[n_m])\cD_q(-i[n_m])\cD_{\tq}(-i[n_m])$ 
can be quite readily evaluated from (\ref{DW}), (\ref{Dq}) and (\ref{Dtq}) and using symmetries of the integrals as
\begin{equation}
\log\cD_W(-i[n_m])
=\frac{2[n_m]}{\pi }\int^{\infty}_{0} \frac{dt}{\cosh t}\left[\log\left(1-e^{-4\pi
R|a|\cosh t+i2\theta_e}\right)
+\log\left(1-e^{-4\pi R|a|\cosh t-i 2\theta_e}\right)\right]\,,
\label{sadcD2}
\end{equation}
\begin{equation}
\begin{aligned}
\log\cD_q(-i[n_m])=
-\sum_{i=1}^{N_f}\frac{[n_m]}{2\pi }\int^{\infty}_{0} \frac{dt}{\cosh t}\left[\log\left(1-e^{-2\pi
R|a+{m_i}|\cosh t+i(\theta_e+\psi_i)}\right) \right. \\ \left.
+\log\left(1-e^{-2\pi R|a+m_i|\cosh t-i(\theta_e+\psi_i)}\right)\right]\,,
\label{sadcDq}
\end{aligned}
\end{equation}
\begin{equation}
\begin{aligned}
\log\cD_{\tq}(-i[n_m])=
-\sum_{i=1}^{N_f}\frac{[n_m]}{2\pi }\int^{\infty}_{0} \frac{dt}{\cosh t}\left[\log\left(1-e^{-2\pi
R|a-{m_i}|\cosh t+i(\theta_e-\psi_i)}\right) \right. \\ \left.
+\log\left(1-e^{-2\pi R|a-{m_i}|\cosh t-i(\theta_e-\psi_i)}\right)\right]\,.
\label{sadcDtq}
\end{aligned}
\end{equation}
In deriving (\ref{sadcDq}) and (\ref{sadcDtq}), we have also used the
residual flavor $U(1)^{N_f}$ symmetries to set ${\rm Im}(m_i/a)=0$,
which is also necessary to ensure they yield real values.  In
\cite{CDP}, it was shown that (\ref{sadcD2}) precisely corresponds to the
ratio of one-loop determinants corresponding to non-zero mode
fluctuations around a monopole in the pure $SU(2)$ case.  Later we
will perform similar computations to demonstrate that (\ref{sadcDq})
and (\ref{sadcDtq}) indeed correspond to fundamental hypermultiplet
non-zero mode fluctuations around a monopole.
\paragraph{}
In this paper, we shall focus on the leading one-instanton correction
to the moduli space $\cM$. This restricts us to the subsector
$k=n_m=1$ in the series (\ref{wNP}), and we can extract it from the
$\zeta$-independent part of $\omega_\ga(\z)$ as:
\begin{eqnarray}\label{explicitw3inst}
\w^{\rm inst.}_{3}
&=&\sum_{\ga=(\gae,1)}\Omega(\ga,u)
\cJ_{(\ga,1)}\left((2\pi R) dZ_{\ga}\wedge d\bar{Z}_{\ga}
+i|Z_{\ga}|d\th_{\ga}\wedge\left(\frac{dZ_{\ga}}{Z_{\ga}}-\frac{d\bar{Z}_{\ga}}{\bar{Z}_{\ga}}\right)\right)\,.
\end{eqnarray}
Using the definitions $Z_\ga=n_e a+n_m a_D+\sum_{i=1}^{N_f}s_i {m_i}$ and $\tau_{\rm eff}=\frac{d a_D}{d a}$,
we can write out the $g_{a\ba}$ component from above:
\begin{equation}\label{gaainst1}
g_{a\bar{a}}^{\rm
inst.}=\frac{\sqrt{R}}{8\pi}\sum_{\ga=(\gae,1)}\Omega(\ga,u)\frac{
| n_e+\tau_{\rm eff}|^2}{|Z_{\ga}|^{1/2}}\cD(-i)\exp\left[-2\pi R|Z_{\ga}|+i\left(\th_{\ga}+\sum_{i=1}^{N_f} s_i \psi_i\right)\right]\,.
\end{equation}
Other metric components $g^{\rm inst.}_{a\bar{z}}, g^{\rm inst.}_{\ba z}$ which
are suppressed by $g^2_\eff$, can also be readily extracted from
(\ref{explicitw3inst}).
However for our later comparison with semi-classical computation, 
it is sufficient to expand the metric (\ref{gaainst1}) at weak coupling as
\begin{eqnarray}
g_{a\bar{a}}^{\rm inst.}&\approx
&\frac{\sqrt{R}}{8\pi}\left(\frac{8\pi}{g^2_\eff}\right)^{3/2}
\sum_{\ga=(\gae,1)}\Omega(\ga,u)\frac{\cD(-i)}{|a|^{1/2}}\exp\left( -S_{\rm
Mon}-S_{\varphi}^{(\gae)}\right)\,,\label{gaaApprox}\\
S_{\rm Mon.}&=&(2\pi R)\frac{8\pi}{g^2_\eff}|a|-i \theta_m\,,\label{Smon}\\
S_{\varphi}^{(\gae)}&=&\frac{
1}{2}\frac{(2\pi R) |a|}{8\pi/g_{\rm eff}^2}\left[\left(\gae+\frac{\Theta_\eff}{\pi}\right)|a|+\sum_{i=1}^{N_f} s_i {|m_i|}\right]^2-i\left[\left
(\gae+\frac{\Theta_\eff}{\pi}\right)\theta_e+\sum_{i=1}^{N_f}s_i\psi_i\right]\,.\label{Schi}
\end{eqnarray}
The term $S_{\rm Mon}$ here is essentially the Euclidean action of a
magnetic monopole dimensionally reduced on $\mathbb{R}^{3}\times
S^{1}$.  The remaining terms in $S_{\varphi}^{(\gae)}$ are the leading
contributions from the dyon electric charge. When $m_i=\tm_i=0$, the
shifted electric charge $n_e+\frac{\Theta_{\rm eff}}{\pi}$ in the
quadratic term reflects the combination of global $U(1)$ rotation in the
monopole moduli space \cite{tw} and the well-known Witten effect
\cite{wt}.  The linear $\Theta_{\rm eff}$ shift in the second term
of $S_{\varphi}^{(n_e)}$ {should be introduced to account for the shift} $\theta_m\to
\theta_m+\frac{\Theta_{\rm eff}}{\pi}\theta_e$, and as explained in
detail in \cite{GMN1, CDP}, this is required to ensure the single-valuedness of
$\theta_m$ near the singularity at infinity; {$\sigma(\gamma)$ is absorbed by the global definition of $\theta_m$ (cf.\ eq.\ (4.16b) in \cite{GMN1})} \footnote{Notice that despite the coefficient of the beta function changes for the flavor case,
we can still use $\tau_{\rm eff}=\frac{d a_D}{da}$ to re-express the shift.}.  
We shall postpone the
discussion for the $m_i\,, \tm_i \neq 0$ cases to the next section
when we consider first principle semi-classical computations.
\paragraph{}
To facilitate the explicit comparison with the four fermion vertex arising from the semi-classical computation, we also need to compute the non-perturbative instanton corrections to the Riemann tensor. 
This can be readily computed from (\ref{gaaApprox}), up to the permutation symmetries, 
the leading components in $g_{\rm eff}^2$ expansions are:
\begin{equation}\label{expRiemanntensor}
R_{a\bar{z}z\ba}=R_{a\ba z\bar{z}}=-\frac{1}{4} g_{a\ba}^{\rm inst.}\,.
\end{equation}
We can now extract the non-perturbative corrections to the low-energy
effective action for the three-dimensional supersymmetric sigma model 
up to most two derivatives and four fermions terms:
\begin{equation}\label{3DSeff}
S_{\rm eff}^{\rm (3D)}=\frac{1}{4} \int d^3 x \left(g_{ij}(X)\left[\partial_\mu
X^i\partial^{\mu} \bar{X}^{j}+i\bar{\Omega}^{i}  \ssl{D}\Omega^j \right]
+\frac{1}{6} R_{ijkl}(\bar{\Omega}^{i}\cdot\Omega^k)(\bar{\Omega}^{j}\cdot
\Omega^l)\right)\,,
\end{equation}
where $\{X^i\}$ are four bosonic scalar fields and
$\{\Omega_i^\alpha\}$ are their Majorana fermionic superpartners.
We followed the conversion procedures in \cite{CDP}
between $(X^i, \Omega^i)$, the bosonic $(a,z)$, and the fermionic $(\lambda,\psi)$ fields 
in the dimensionally reduced actions (\ref{bosonic 3D action}) and (\ref{fermionic 3D action})
\footnote{Note that the $g_{\rm eff}$ in \cite{CDP} differs from ours by $\sqrt{2}$, 
due to the different normalization for the complex gauge coupling $\tau_{\rm eff}$ (\ref{Deftau}) for the flavor case.}.
After taking this into account, we obtain that the four fermion vertex due to single instanton sector is given by 
\begin{equation}\label{4fermi}
S_{\rm 4F}=\frac{2^7\pi}{R|a|^{1/2}} \left(\frac{2\pi
R}{g^2_{\rm eff}}\right)^{7/2}\cD(-i)
\exp\left[-S_{\rm Mon}\right]
\sum_{\gae\in\bbZ}\Omega(\gamma,u)\exp\left[-S_{\varphi}^{(\gae)}\right]\int d^3
x(\psi\cdot\bar\psi)(\lambda\cdot\bar\lambda)\,.
\end{equation}
We shall next verify this term in the effective action via a
direct semiclassical calculation.

\section{Semiclassical Instanton Calculation with Matters}
\paragraph{}
In this Section we will perform first-principle computation for the monopole and dyon contributions to
the low energy effective action for our $\cN=2$ $SU(2)$ theory with $N_f$ fundamental hypermultiplets compactified on $\mathbb{R}^3\times S^1$. 
We shall focus on the half-BPS states with $\gam=1$ and arbitrary
electric charges $\gae \in \bbZ$, preserving four out of eight supersymmetries,
and their contributions to the four fermion correlation function.
A closely related computation for the compactified pure $\cN=2$ $SU(2)$ theory has recently been done in \cite{CDP}, 
we shall therefore refer readers to it for some of the technical details, 
in following we shall instead highlight the essential modifications due to the fundamental hypermultiplets.
\paragraph{}
The main object of interests here is the four fermion correlation function of the form: 
\begin{equation}
\cG_{4}(\by_1,\by_2,\by_3,\by_4)=\langle{\prod_{A=1}^2
\rho^{ ~ A}_1(\by_{2A-1})\rho^{~
A}_2(\by_{2A})}\rangle\,,
\label{corr}
\end{equation}
where $\rho_{1,2}^A$ are Weyl fermions of purely left handed four dimensional chirality. 
We would like to evaluate (\ref{corr}) in the monopole background, and from its large distance behavior
such that $\rho_{1,2}^A$ take their zero mode values, 
we can extract the coefficient for the four fermion vertex $\sim (\bar{\rho}^{1}\cdot\bar{\rho}^{1})\,\,
(\bar{\rho}^{2}\cdot\bar{\rho}^{2})$ in the low-energy effective
action. 
Let us comment on the subtlety explained in \cite{CDP} and \cite{Dorey2000A} relating $\rho^{A}_{1,2}$ and the 3d fermions $\psi_{1,2}$ and $\lambda_{1,2}$ in the previous section. $\rho^{A}_{1,2}$ in (\ref{corr}) are in fact chiral fermions in a compactified auxiliary four dimensional theory, 
where the BPS equation can be identified with the dimensionally reduced self-duality equation, and the monopole preserves supercharges of the same four dimensional chirality.  
The fermions $\rho^A_{1,2}$ of the auxiliary theory are related
to the original four-dimensional Weyl fermions $\psi_{1,2}$ by an $SO(3)$
$R$-symmetry rotation which mixes left and right-handed
chiralities but preserves the normalisation of the four-fermion
vertex in the effective Lagrangian.
In a vacuum where $\theta_{e}=0$, the zero modes of a monopole
are chirally symmetric in the original four-dimensional theory,
and the explicit relation takes the form 
\begin{eqnarray}
(\bar{\rho}^{1}\cdot\bar{\rho}^{1})(\bar{\rho}^{2}\cdot\bar{\rho}^{2}) & = &
(\psi\cdot\bar\psi)(\lambda\cdot\bar\lambda)\,.
\label{transfmn}
\end{eqnarray}
\paragraph{}
Let us now consider the zero modes in monopole background, the Callias
index theorem \cite{Callias} tells us there are $4n_m$ real bosonic
zero modes for monopole configuration of charge $n_m$.  In our case
therefore there are four bosonic zero modes for $n_m=1$ monopole:
$X^{1,2,3}$ parametrising its center position in ${\mathbb R}^3$ and a
global $U(1)$ charge angle $\varphi$, the bosonic moduli space is
therefore ${\mathbb R}^3 \times S^1_{\varphi}$.  For general $n_m$,
the remaining $4(n_m-1)$ bosonic zero modes parametrise the relative
moduli space.  There are also $4n_m$ adjoint fermionic zero modes for
our theories with eight supercharges, in particular four of them are
generated by the action of the four broken supersymmetries on the
$(X^{1,2,3},\varphi)$ and we can denote the corresponding collective
fermionic coordinates $\xi^A_{1,2}$, these fermionic zero modes are
protected from lifting by supersymmetries.  The large distance limit
$|\by-X| \gg |a|^{-1}$ of $\rho^{A}_{\alpha}$ is then given by
\begin{equation}\label{largeDFzeromode}
\rho_\alpha^{{\rm (LD)} ~ A}(\by)=16\pi (S_F(\by-X))^\beta_\alpha
\xi^{A}_\beta\,,
\end{equation}
where $S_F(x)=\gamma^\mu x_\mu/(4\pi |x|^2)$ is the three dimensional Dirac propagator.
The remaining $4(n_m-1)$ fermionic zero modes are essentially the supersymmetric partners of the $4(n_m-1)$ bosonic coordinates on the relative moduli space.
\paragraph{}
Now when we include additional $N_f$ fundamental hypermultiplets $(q_i,\tq_i)$ with masses $m_i$ and $\tm_i$, they can also contribute additional zero modes in the monopole background, hence it is necessary to perform an index computation to count their numbers. To do so, we follow 
\cite{Weinberg, Kaul, DTV} to define the following four dimensional fluctuation operators for the massive fundamental hypermultiplets in the monopole background:
\begin{eqnarray}
&&\Delta_+^{q_i}(m_i)=-\vec{D}^2+\left|a+{m_i}\right|^2\,,\label{Deltaqplus}\\
&&\Delta_-^{q_i}(m_i)=-\vec{D}^2+2 \epsilon_{ijk}\sigma_i F_{jk}^{\rm
Mon.}+\left| a+{m_i}\right|^2\,.\label{Deltaqminus}
\end{eqnarray}
Here the three dimensional covariant derivative $\vec{D}=\vec{\partial}+ i \vec{A}^{\rm
Mon.}$ is with respect to background static monopole in $A_0=0$ gauge.     
We can define similar operators $\Delta_{\pm}^{\tq_i}(m_i)$ for $\tq_i$ with $|a+m_i|\to |a-m_i|$.
The number of the (complex) hypermultiplet zero modes coming from $q_{i}$ and $\tq_{i}$ then comes from the $\mu^2\to 0$ limit of the regularized trace:
\begin{eqnarray}\label{hyperInd}
\cI_{\rm H}(m_i)={\rm Tr}_i\left[\frac{\mu^2}{\Delta_-^{q_i}(m_i)+\mu^2}
-\frac{\mu^2}{\Delta_+^{q_i}(m_i)+\mu^2}\right]
+{\rm Tr}_i\left[\frac{\mu^2}{\Delta_-^{\tq_i}(m_i)+\mu^2}
-\frac{\mu^2}{\Delta_+^{\tq_i}(m_i)+\mu^2}\right]\,,
\end{eqnarray}
where ${\rm Tr}_i$ indicates summing over the flavor indices and normalizable states.
The trace in (\ref{hyperInd}) can be evaluated analogously following the steps in \cite{Weinberg, WeinbergYi} for monopole of charge $n_m$, the result is: 
\begin{eqnarray}\label{DefIq}
\cI_{\rm H}(m_i)&=&\sum_{i=1}^{N_f}\frac{n_m}{2}\left[
\frac{|a|+|m_i|}{\left(\left( |a|+|m_i|\right)^2+\mu^2\right)^{1/2}}
+\frac{|a|-|m_i|}{\left(\left( |a|-|m_i|\right)^2+\mu^2\right)^{1/2}}\right]\,,
\end{eqnarray}
where in writing out $\cI_{\rm H}(m_i)$ we have also used the fact ${\rm Im}(a/m_i)=0$.
In the $\mu^2 \to 0$ limit, we have $\cI_H(m_i)\to \frac{n_m N_f}{2}[{\rm sign}(|a|+|m_i|)+{\rm sign}(|a|-|m_i|)]$. 
Since in the weak coupling we expect $|a|\gg |m_i|$, there are $2 n_m N_f$ additional real zero modes appearing.
As discussed in \cite{GaHa, WeinbergYi, ManSch, GKLY}, these additional hypermultiplet zero modes facilitate a natural $\cO(n_m)$ bundle over the 
$n_m$ monopole moduli space, and they are required to form bound states with the BPS monopoles/dyons for them to transform under the flavor symmetry group \cite{SW2}. In our computation of single monopole $n_m=1$, the $\cO(1)$ index bundle ${\rm Ind}_1 =R^3\times \text{M\"ob}$, where $\text{M\"ob}$ is the M\"obius bundle over $S^1$ of the monopole moduli space. This bundle is obviously flat with vanishing curvature, 
however the non-trivial twisting comes from the fact that the $2\pi$ global rotation about the $S^1$ acts as non-trivial element of the center of $SU(2)$ gauge group \cite{GaHa}. We shall return to this point shortly in the following discussions.

\paragraph{}
Having discussed the zero modes, 
the semiclassical dynamics for a single monopole of mass $M=8\pi |a|/g^2$
can be described by supersymmetric quantum mechanics on its moduli space \cite{Gauntlett1993}. 
The collective coordinate Lagrangian including the hypermultiplet zero modes takes the form \cite{GaHa, GKLY}:
\begin{equation}\label{MonLQM}
L_{QM}=L_X+L_{\varphi}+L_{\xi}+L_{\eta}\,.
\end{equation}
Here the bosonic Lagrangians are $L_X=\frac{M}{2}|\dot{\vec{X}}|^{2}$ and 
$L_{\varphi}=\frac{1}{2}\frac{M}{|a|^2}(\dot{\varphi})^2$,
where the dot denotes the derivative with respect to Euclidean time $x^4$,
and $\vec{X}$ is the position of the monopole in ${\mathbb{R} }^3$. 
The combination $\frac{M}{|a|^2}$ is the moment of inertia of a monopole with
respect to global gauge rotation, $L_{\varphi}$ describes a free particle of
mass $\frac{M}{|a|^2}$ moving along $S_{\varphi}^{1}$ with $\varphi \in
[0,2\pi]$.
The bosonic degrees of freedom are supersymmetrized by the adjoint fermionic collective coordinates $\xi^A_{\alpha}\,,~A,\alpha=1,2$
with the free Lagrangian $L_{\xi}=\frac{M}{2}\xi^A_{\alpha}\dot{\xi}^{\alpha}_A$.
The $2N_f$ real hypermultiplet collective coordinates $\eta^i$ are encoded in the Lagrangian $L_\eta=\frac{1}{2}( \eta_i {\cD}_{x^4}\eta^i+m \eta^2)$,
where $\cD_{x_4}$ is the covariant derivative with respect to the connection on the index bundle, and we have also included the complex mass term.   
We can now write down the large distance behavior of
the four fermion correlation function: 
\begin{equation}
\begin{aligned}
\cG_{4}&(\by_1, \by_2, \by_3, \by_4)
= \int \frac{[d^3 X(x^4)]}{(2\pi)^{3/2}}\frac{[d\varphi(x^4)]}{(2\pi)^{1/2}}[d^4 \xi(x^4)][d^{2N_f} \eta(x^4)]\cR
\\
&\times \prod_{A=1}^2
\rho^{{\rm(LD)} ~ A}_1(\by_{2A-1})\rho^{{\rm(LD)} ~
A}_2(\by_{2A})
\exp\left[-\int^{2\pi R}_0 dx^4 L_{QM}\right]
\exp\left[-\frac{16\pi^2 R |a|}{g^2}+i\th_m\right]\,,
\label{G4}
\end{aligned}
\end{equation}
where the large distance fermionic zero modes $\rho^{{\rm(LD)} ~
  A}_{1,2}$ are as given in (\ref{largeDFzeromode}) \footnote{The
  prefactor of $1/(2\pi)^{2}$ arises from the Jacobian for the change
  of variables from bosonic fields to the four bosonic collective
  coordinates and can be traced to the same factor in the standard
  formula \cite{Ber} given as eq.\ (114) in \cite{Dorey1997}.}.  The
integration measure consists of bosonic $[d^3 X][d\varphi]$
and fermionic $[d^4\xi]$ collective coordinates measures.  
In addtion we also need to integrate over $2N_f$ hypermultiplet collective coordinates $[d^{2N_f}\eta(x^4)]$. 
The one-loop determinant
$\cR$ encoding the non-zero mode fluctuations from both vector and
hypermultiplets, will be evaluated shortly.  
The various terms
are all weighted by the monopole effective action $\exp[-\int^{2\pi
  R}_0 dx^4 L_{QM}-S_{\rm Mon.}]$ given in (\ref{MonLQM}), after
Euclideanizing the time direction and compactifying on $S^1$.  
\paragraph{}
As the fermionic insertion $\rho^A(\by)$ (\ref{largeDFzeromode}) only depends on the three spatial coordinates $\vec{X}$ 
and the adjoint fermionic zero modes $\xi^A_{\alpha}$, we can split the four fermion correlation function as
$\cG_4=\cG_4^{\rm COM}\times {\mathcal Z} \times {\mathcal R}$, where:
\begin{equation}
\begin{aligned}
\cG^{\rm COM}_4(\by_1,\by_2,\by_3,\by_4)
&= \int \frac{[d^3 X(x^4)][d^4 \xi(x^4)]}{(2\pi)^{3/2}}\prod_{A=1}^2
\rho^{{\rm(LD)} ~ A}_1(\by_{2A-1})\rho^{{\rm(LD)} ~
A}_2(\by_{2A}) \\
&\times \exp\left[-\int^{2\pi R}_0 dx^4 \left(L_{X}+L_{\xi}\right)-S_{\rm Mon.}\right]\,,
\label{G4COM}
\end{aligned}
\end{equation}
\begin{equation}
{\mathcal Z}=\int\frac{d\varphi(x^4)}{(2\pi)^{1/2}}[d^{2N_f}\eta(x^4)]\exp\left[-\int^{2\pi R}_{0} dx^4 \left(L_{\varphi}+L_{\eta}\right)\right]\,.\label{PartZ}
\end{equation}  
To evaluate $\cG^{\rm COM}_4$, we impose usual periodic boundary condition
$\vec{X}(x^4)=\vec{X}(x^4+2\pi R)$ and
$\xi^A_{\alpha}(x^4)=\xi^A_{\alpha}(x^4+2\pi R)$ to ensure
supersymmetry is preserved.
For the bosonic and fermionic collective coordinate integration measures,
$\int [d^3 X(x^4)] \exp[\int^{2\pi R}_{0} dx^4 L_{X}]$ and $\int [d^4\xi(x^4)]\exp[\int^{2\pi R}_{0} dx^4 L_{\xi}]$,
the path integrals are dominated by the constant classical paths due to the periodic boundary condition. 
Integrating over the classical paths, we can readily obtain (see \cite{CDP} for more details):
\begin{eqnarray}
\int [d^3 X(x^4)][d^4\xi(x^4)]\exp\left[-\int^{2\pi R}_0 dx^4 \left(L_X+L_\xi\right)\right]
=\int d^3 X \int d^4\xi \left[\sqrt{\frac{M}{2\pi (2\pi
R)}}\right]^{-1}\,,\label{measures}
\end{eqnarray}
where we have also used $\vec{X}$ and $\xi^{A}_\alpha$ in the integrations 
to denote the classical values of the bosonic and fermionic zero modes respectively. 
\paragraph{}
To evaluate $\cZ$ next, 
here is a good place to recall an alternative interpretation for the four fermion correlator $\cG_4(\by_1,\by_2,\by_4,\by_4)$ (\ref{corr}) in the compactfied theory, following \cite{Dorey2000A, DHK2001}.
That is, we can instead work in the Hamiltonian formalism, and regard it as a generalization of Witten index:
\begin{equation}
\langle{\prod_{A=1}^2
\rho^{ ~ A}_1(\by_{2A-1})\rho^{~
A}_2(\by_{2A})}\rangle={\rm Tr}\left(\prod_{A=1}^2
\rho^{ ~ A}_1(\by_{2A-1})\rho^{~
A}_2(\by_{2A})(-1)^{F}\exp\left[-2\pi R\, H_{QM}-S_{\rm Mon.}\right]\right)\label{corrPart}\,.
\end{equation}
Here $H_{QM}$ is the Hamiltonian for the collective coordinates Lagragian $L_{QM}$, 
the trace ${\rm Tr}$ sums over the BPS states which will be discussed immediately below, 
each contributes with the exponential suppression factor $\exp\left[-S_{\rm Mon.}\right]$.
In particular, the interpretation above allows us to re-express the factor $\cZ$ as:
\begin{equation}
\cZ={\rm Tr}_i\left( (-1)^F {\mathbb P}_{n_e} \exp\left[-2\pi R \left(H_{\varphi}+H_{\eta}\right)\right]\right)\,,\label{ZPart}
\end{equation}
where ${\mathbb P}_{n_e}$ is a projector which depends on the electric charge of the state, and the subscript $i$ indicates that we also need to sum over the representation under the flavor group. 
In the semi-classical quantization of monopole quantum mechanics, 
the wave function of any BPS states on the one-monopole moduli space can be decomposed schematically 
into a tensor product $|\Psi_{\rm BPS} \rangle =f(\vec{X},\varphi) |\xi\rangle \otimes |\eta_i \rangle$. 
$f(\vec{X},\varphi)$ is the bosonic part and a function of collective coordinate $\{\vec{X},\varphi\}$, 
$|\xi\rangle$ comes from their SUSY partners; the remaining $|\eta_i\rangle$ ensures the BPS states transform 
as spinors under the $SO(2N_f)$ group (massless) or are charged under the residual $U(1)^{N_f}$ (massive).
\paragraph{}
Recall that without the flavors, a $2\pi$ global rotation about the $S^1$ in the moduli space, with generator $Q$, 
leaves the monopole wave function invariant, i.e. $e^{2\pi i Q}=1$. 
This gives rise to the whole tower of quantized electric charges $n_e \in {\mathbb Z}$ \cite{tw},  
which we can identify with the quantized conjugate momentum $P_{\varphi}=\frac{M}{|a|^2} \dot{\varphi}=n_e$.
The corresponding Hamiltonian is $H_\varphi=\frac{1}{2}\frac{|a|^2}{M} n_e^2$, 
and the trace in (\ref{ZPart}) essentially sums over these BPS dyons in the theory. 
In the presence of flavors, there is a key modification in the above discussion \cite{SW2} (see also \cite{Harvey1996} for nice discussion). 
Now the $2\pi$ rotation about $S^1$ does not give identity but 
a topologically non-trivial gauge transformation, whose eigenvalue is given by $e^{i\pi Q}=e^{i\Theta}(-1)^H$ 
\footnote{The factor of 2 for the generator $Q$ is to ensure that the fundamental quarks have charge $\pm1$ and W-bosons have charge $\pm 2$.}, 
where $(-1)^H$ is the center of $SU(2)$ gauge group, and $\Theta$ is the Witten angle. 
If we set the electric charge $Q=n_e +\frac{\Theta}{\pi}$ with $n_e \in {\mathbb Z}$, 
this yields that the states of odd $n_e$  have chirality operator $(-1)^H$ odd, and the states of even $n_e$ have $(-1)^H$ even. 
This also implies that when $m_i=0$, $(-1)^H$ plays the analog of $\gamma^5$ in the $SO(2N_f)$ Clifford Algebra facilitated by the collective coordiantes $\eta^i$ \cite{Harvey1996}, and we can form the projection operators ${\mathbb P}_{n_e}=\frac{1}{2}(1\pm (-1)^H)$ (as the one inserted in (\ref{ZPart})) to decompose the original reducible $2^{N_f}$ dimensional spinor representation down to two irreducible $2^{N_f-1}$ representations with definite electric charge $n_e$.
Similar analysis can also be done for $m_i \neq 0$, which decomposes the wave functions into components carrying definite charge under each $U(1)$ of the residual $U(1)^{N_f}$ group, whose value depends on $n_e$.
The extensive discussion above therefore allows us to conclude that
\begin{equation}
\cZ=\sum_{\gae\in \bbZ}\Omega((n_e,1),u)
\exp\left[-\frac{\pi R}{M}\left(\left(\gae+\frac{\Theta}{\pi}\right)|a|+\sum_{i=1}^{N_f}s_i{|m_i|}\right)^2+
i\left(\gae+\frac{\Theta}{\pi}\right)\theta_e+i\sum_{i=1}^{N_f} s_i \psi_i\right]\,.\label{ZPart2}
\end{equation}
In above, for definite $n_e$, with the projector ${\mathbb P}_{n_e}$ imposing the anti-periodic boundary condition,
the degenaracy factor $\Omega((n_e,1),u)$ comes from tracing over the factor $e^{-2\pi R H_\eta}$ over different flavor $|\eta^i\rangle$, 
and it should be identified with $\Omega(\gamma,u)$ appearing in (\ref{gaaApprox}).
Here we have also included the shift of electric charge due to mass terms $(m_i,\tilde{m}_i)$: this can be motivated from our earlier choice ${\rm Im}(a/m_i)=0$, and the $S^1$ rotation is now a linear combination of the global $U(1)$ within the gauge $SU(2)$ and the residual $U(1)^{N_f}$ flavor group.
A further phase $n_e\theta_e+\sum_{i=1}^{N_f} s_i \psi_i$ in the classical action arises from the surface terms
coupling to electromagnetic charge and the Wilson line for the flavor group. 
In summary, we note that this matches the corresponding sum appearing in
the GMN prediction (\ref{gaaApprox}) up to a replacement of the bare
coupling and the vacuum angle by their one-loop renormalised counterparts.

\paragraph{}
To complete the semiclassical integration measure, 
for our cases with $N_f$ fundamental hypermultiplets on $\mathbb{R}^3\times S^1$,
it is necessary to evaluate the one loop determinant $\cR$ 
accounting for the non-zero mode fluctuations in the monopole background.
We can decompose $\cR$ into three components:
\begin{equation}\label{3cR}
\log\cR=\log \cR_W+\log \cR_{q}+\log\cR_{\tq}\,,
\end{equation}
where $\log \cR_W$ is the W-boson contribution, while $\log \cR_q$ and $\log \cR_{\tq}$
are the additional hypermultiplet contributions.
In \cite{CDP}, $\cR_W$ was explicitly computed using Kaul's earlier result for the density of states of   
the fluctuations in the monopole background \cite{Kaul},  
we shall follow similar steps here to calculate $\cR_{q}$ and $\cR_{\tq}$.  
\paragraph{}
Let us begin with $\cR_q$, the computation for $\cR_{\tq}$ can be carried out analogously.
Essentially, all the spatial fluctuations of the hypermultiplets can be simply encoded by  
the non-zero eigenfunctions of the fluctuation operators $\Delta^{q_i}_{\pm}$ (\ref{Deltaqplus}) and (\ref{Deltaqminus}).
To further include the fluctuations along $S^1: x^4 \sim x^4+2\pi R$, 
we define: 
\begin{eqnarray}\label{Deltapm}
{\bD}_{\pm}^{q_i}={\Delta}_{\pm}^{q_i}+\left(\frac{\partial}{\partial
x^4} \right)^2\,,
\end{eqnarray}
where the derivative with respect to $x_{4}$ take account of the
Fourier modes of each fluctuation mode on $S^1$, and we have first set $\theta_e=\psi_i=0$ to simplify the discussion.  
Summing over $N_f$ different flavors, 
the one-loop contribution determinant associated with the $q_i$ fluctuations is given by the ratio: 
\begin{equation}\label{Def1loop}
\cR_q=\prod_{i=1}^{N_f}\left[\frac{\det (\bD_{+}^{q_i})}{\det (\bD_{-}^{q_i}) }\right]^{-1/2}\,.
\end{equation}
To evaluate (\ref{Def1loop}), we
can decompose any eigenfunctions of $\bD_{\pm}^{q_i}$ as
$\Phi_{\pm}(\vec{x},x^4)=\phi_{\pm}(\vec{x})f_{\pm}(x^4)$ by $S^1$ translational invariance,
where $\phi_{\pm}(\vec{x})$ are eigenfunctions of $\Delta_{\pm}^{q_i}$ with eigenvalues $\lambda_{\pm}^2$ respectively.
While $f_{\pm}(x^4)$ along the $S^1$ take the plane-wave form
$f_{\pm}(x^4)\sim e^{i\varpi_{\pm} x^4}$.
In a supersymmetric gauge theory,
the total number of non-zero eigenvalues for both bosonic and
fermionic fluctuations are exactly equal, this usually implies their contributions 
to (\ref{Def1loop})
completely cancel and $\cR_{q}=1$.  
However, the spectra of $\Delta_{\pm}^{q_i}$ contain both
normalizable bound states and continuous scattering states, and
the precise cancellation requires exact densities of states for
bosonic and fermionic eigenvalues.
As discovered by \cite{Kaul}, this is not the case in the monopole
background, resulting in non-trivial quantum corrections to the monopole mass.
In our case, the operators $\bD_{\pm}^{q_i}$ on $\mathbb{R}^3\times S^1$ also inherit 
such subtle effect from $\Delta_{\pm}^{q_i}$, giving non-trivial $\cR_q$.  
\paragraph{}
We can rewrite $\cR_q$ (\ref{Def1loop}) as the following integral expression:
\begin{eqnarray}
\cR_q 
&=&\exp\left[-\frac{1}{2}\sum_{i=1}^{N_f}\int_{|a+{m_i}|}^{\infty} d\lambda
\delta
\rho_i(\lambda) \log\left[ \cK_i(\lambda,2\pi R)\right]\right]\,,\label{Int1loop}\\
\cK_i(\lambda,2\pi R)&=& \det{}_{x^4}\left[\left(\frac{\partial}{\partial
x^4}\right)^2+\lambda^2\right]\,,\label{DefKlambdaR}
\end{eqnarray}
where we have used the identity $\log {\rm det}({\rm M})={\rm Tr}\log({\rm M})$.
The quantity
$\delta\rho_i(\lambda)=\rho_{+,i}(\lambda)-\rho_{-,i}(\lambda)$
is the difference between densities
of eigenvalues of the operators $\Delta_{+}^{q_i}$ and $\Delta_{-}^{q_i}$.
This can be worked out using index theorem following 
\cite{Kaul} and recycling our earlier computation for the hypermultiplet index $\cI_{\rm H}(\vec{m})$, 
yielding
\begin{equation}\label{deltarho}
d\lambda \delta \rho_i(\lambda)=-\frac{\left|a+{m_i}\right| d\lambda^2}{2\pi
\lambda^2\sqrt{\lambda^2-\left|a+{m_i}\right|^2}} \,.
\end{equation}
As noted in \cite{CDP}, the integration kernel $\cK_i(\lambda,2\pi R)$
is precisely the partition
function of
harmonic oscillator with frequency $\varpi_{i}=\lambda$ at inverse temperature
$\beta=2\pi R$.
$\theta_e$ and $\psi_i$,
which are non-vanishing VEV for the $x^4$-component of the respective gauge fields,
can be minimally coupled to the operators $\bD_{\pm}^{q_i}$ given in (\ref{Deltapm}):
\begin{eqnarray}
\frac{\partial}{\partial x_{4}} & \rightarrow &
\frac{\partial}{\partial x_{4}}\pm\frac{1}{2\pi R}(\theta_e+ \psi_i)\,.
\label{Kshift}
\end{eqnarray}
This is equivalent to introducing a chemical potential to the aforementioned harmonic oscillator system and shifting its
frequencies to the complex values $\varpi_{i}=\lambda\mp
\frac{i}{2\pi R}(\theta_e+ \psi_i)$.
The $\pm$ signs in (\ref{Kshift}) take into account electric $n_e$ and flavor $s_i$ charges for the fluctuations associated with $q_i^{\pm}$.  
Summing over both contributions, we find $\cK_i=\cK_{+,i}\cK_{-,i}$ where
\begin{equation}\label{cKpm}
\cK_{\pm, i}(\lambda,\theta_e, 2\pi R)^{-1}= \frac{\exp[-\pi R \lambda\pm
i(\theta_e+\psi_i)/2]}{1-\exp[-2\pi R \lambda \pm i (\theta_e+ \psi_i)]}\,.
\end{equation}
\paragraph{}
Substituting (\ref{cKpm}) and (\ref{deltarho}) into (\ref{Int1loop}), and changing variable $\lambda=\left|a+{m_i}\right|\cosh t$,
for the one-loop determinant $\cR_q$ we get:
\begin{equation}\label{cR3}
\begin{aligned}
\log \cR_q &= \sum_{i=1}^{N_f} R\left|a+{m_i}\right|\cosh^{-1} \frac{|\Lambda_{\rm
UV}|}{\left|a+{m_i}\right|}
-\frac{1}{2\pi }\sum_{i=1}^{N_f}\int^{\infty}_{0} \frac{dt}{\cosh t}\left[\log\left(1-e^{-2\pi
R\left|a+{m_i}\right|\cosh t+i (\theta_e+\psi_i)}\right) \right. \\
&+\left.\log\left(1-e^{-2\pi
R\left|a+{m_i}\right|\cosh t-i (\theta_e+\psi_i)}\right)\right]\,,
\end{aligned}
\end{equation}
where we have evaluated the integral over the eigenvalues with a UV
cut-off $\Lambda_{\rm UV}$. 
We immediately recognise that the  
integral here is precisely the same one in the
definition (\ref{sadcDq}) of $\log \cD_q(-i)$ in the
semiclassical expansion of the GMN result.
Furthermore, the $\Lambda_{\rm UV}$-dependent term should be cancelled 
by the corresponding counter-term in the coupling constant renormalisation.
The net effect is the contribution, along with $\tq_i$ and W-boson fluctuations,
to the finite one-loop renormalisation of 
the complex gauge coupling $\tau$, which gives $\tau_{\rm eff}$.
We can follow similar steps to evaluate $\cR_{\tq}$ by changing $(m_i,\tm_i)$ into $-(m_i,\tm_i)$,
while for the W-boson contribution $\cR_W$, we can recover the result in \cite{CDP} by setting $m_i=\tm_i=0$ 
and replacing $(|a|, \theta_e)$ in the operators $\bD^{q_i}_{\pm}$ by $2(|a|,\theta_e)$, 
as the non-zero mode fluctuations for W-bosons carry electric charges of $\pm 2$. 
A further $1/(2\pi R)^2$ also needs to be introduced for $\cR_W$ to account for removal of zero modes in the functional determinant 
as well as matching with the three dimensional limit computed in \cite{Dorey1997} (see \cite{CDP}).  
The resultant $\cR_W$ and $\cR_{\tq}$ again match with (\ref{sadcD2}) and (\ref{sadcDtq}) up to $\Lambda_{\rm UV}$-dependent terms, 
which combine with the $\Lambda_{\rm UV}$-dependent term in $\cR_q$ (\ref{cR3}) to give the renormalised $\tau_{\rm eff}$.
\paragraph{}
Putting all the pieces together and summing over electric charges
$\gae$, we express the large-distance
behaviour of the four-fermion correlation function
$\cG_4(\by_1,\by_2,\by_3,\by_4)$ as
\begin{equation}\label{4ptfunction}
\begin{aligned}
\cG_4(\by_1,\by_2,\by_3,\by_4)=\frac{2^{10}\pi}{R|a|^{1/2}}\cD(-i)\left(\frac{2\pi
R}{g_\eff^2}\right)^{-1/2}
\exp\left[-S_{\rm Mon.}\right]
\sum_{\gae\in\bbZ}\Omega((\gae,1),u)\exp\left[-S_{\varphi}^{(\gae)}\right]
\\
\times \int d^3 X \epsilon^{\alpha'\beta'}\epsilon^{\gamma'\delta'}
S_F(\by_1-X)_{\alpha\alpha'}S_F(\by_2-X)_{\beta\beta'}S_F(\by_3-X)_{\gamma\gamma'}S_F(\by_4-X)_{\delta\delta'}
\end{aligned}
\end{equation}
where we have substituted (\ref{largeDFzeromode}), (\ref{measures}) and (\ref{cR3}) into
(\ref{G4}), and the actions $S_{\rm Mon.}$ and $S_{\varphi}^{(\gae)}$ are as given in (\ref{Smon}) and (\ref{Schi}).
For consistency, we should also use
the same renormalised $g^{2}_{\rm eff}(a)$ wherever the coupling appears. 
The resultant four-fermion correlator corresponds to the appearance of the
following four-fermion interaction vertex in the low-energy effective action:
\begin{equation}\label{4fermiS}
S_{\rm 4F}=\frac{2^{7}\pi}{R|a|^{1/2}} \left(\frac{2\pi
R}{g^2_{\rm eff}(a)}\right)^{7/2}\cD(-i)\exp\left[-S_{\rm Mon}\right]
\sum_{\gae\in\bbZ}\Omega((\gae,1),u)\exp\left[-S_{\varphi}^{(\gae)}\right]\int d^3 x
(\psi\cdot\bar\psi)(\lambda\cdot\bar\lambda)\,.
\end{equation}
This exactly matches the prediction obtained from the integral
equations of \cite{GMN1} given in (\ref{4fermi}).

\section{Three Dimensional Limit and Brane Picture}
\label{sec:3d-and-branes}

\subsection*{Recovering Three Dimensional Quantities}
\paragraph{}
In this section we would like to demonstrate 
how some of the physical quantities computed in strict three dimensional limit \cite{DTV,dBHO} 
may be recovered from our earlier results. 
We shall first focus on the one-loop determinants $\cD_q$ and $\cD_{\tq}$, as given in (\ref{Dq}) and (\ref{Dtq}):
it is clear that when $R|a| \to 0$, the integrals can become divergent and it is necessary perform Poisson resummation \footnote{Our convention for Poisson resummation is $\sum_{k=-\infty}^{+\infty}f(k)=\sum_{n=-\infty}^{+\infty}\widehat
f(n)\,, \quad
\widehat f(n)=\int_{-\infty}^{+\infty}f(k)\,e^{-2\pi i
n k}dk$.}
to obtain finite expressions. The three dimensional limit of $\cD_W$ has been explored in \cite{CDP}.
Let us first Taylor expand $\cD_q$ and $\cD_{\tq}$ along the integration contours, similar to (\ref{AW}) and (\ref{VW}),
and resum over the mode number $k$. We obtain:
\begin{eqnarray}\label{PSDq}
\log \cD_{q}(-i)&=&-\frac{1}{2}\sum_{i=1}^{N_f}\sum_{n\in \bbZ}\left(\sinh^{-1}\left(\frac{\left|a+{m_i}\right|}{\frac{n}{R}+\frac{(\theta_e+\psi_i)}{2\pi R}}\right)-\kappa_n R \left|a+{m_i}\right|\right)\nonumber\\
&+& \sum_{i=1}^{N_f} R\left|a+{m_i}\right|\left(\log\frac{\left|\Lambda_{\rm UV}\right|}{\left|a+{m_i}\right|}+1\right)\,,\\
\log \cD_{\tq}(-i)&=&-\frac{1}{2}\sum_{i=1}^{N_f}\sum_{n\in \bbZ}\left(\sinh^{-1}\left(\frac{\left|a-{m_i}\right|}{\frac{n}{R}+\frac{(\theta_e-\psi_i)}{2\pi R}}\right)-\kappa_n R \left|a-{m_i}\right|\right)\nonumber\\
&+& \sum_{i=1}^{N_f} R\left|a-{m_i}\right|\left(\log\frac{\left|\Lambda_{\rm UV}\right|}{\left|a-{m_i}\right|}+1\right)\,,\label{PSDtq}
\end{eqnarray}
where $\kappa_n$ is regularization constant.
We shall now restrict ourselves to the $\sinh^{-1}(\dots)$ terms, as the other terms proportional to $R|a\pm m_i|$ would vanish in three dimensional limit.
Writing them out explicitly in terms of logarithms and exponentiating, the product $\cD_q(-i)\cD_{\tq}(-i)$ gives
\begin{equation}\label{3dimDqDtq}
\prod_{i=1}^{N_f}\prod_{n\in \bbZ}\left(\frac{\sqrt{\left|a+{m_i}\right|^2+\left(\frac{n}{R}+\frac{(\theta_e+\psi_i)}{2\pi R}\right)^2}-\left|a+{m_i}\right|} {\sqrt{\left|a+{m_i}\right|^2+\left(\frac{n}{R}+\frac{(\theta_e+\psi_i)}{2\pi R}\right)^2}+\left|a+{m_i}\right|}\frac{\sqrt{\left|a-{m_i}\right|^2+\left(\frac{n}{R}+\frac{(\theta_e-\psi_i)}{2\pi R}\right)^2}-\left|a-{m_i}\right|}{\sqrt{\left|a-{m_i}\right|^2+\left(\frac{n}{R}+\frac{(\theta_e-\psi_i)}{2\pi R}\right)^2}+\left|a-{m_i}\right|}\right)^{1/4}\,,
\end{equation}
where $n/R$ should be regarded as KK momentum over the $S^1$.
In the $R\to 0$ limit, which should be taken while keeping fixed the combination $\frac{\theta_e\pm \psi_i}{2\pi R}$, 
all $n\neq 0$ terms in the product above simply yield $1$.
In three dimensions, there are enhanced $SU(2)_N$ symmetry under which $({\rm Re}(a),{\rm Im}(a),\frac{\theta_e}{2\pi R})$ 
and $({\rm Re}(m_i),{\rm Im}(m_i),\tm_i)$ transform as vectors (note that $\psi_i=2\pi R \tm_i$). 
We can therefore exchange $\left|a\pm {m_i}\right|$ and $\frac{\theta_e\pm \psi_i}{2\pi R}$ in (\ref{3dimDqDtq}), 
and further rotate into a vacuum $|a|=0$ vacuum,
so that (\ref{3dimDqDtq}) yields the one-loop determinant $R_H$ for hypermultiplets obtained in \cite{DTV}. 
\paragraph{}
We can similarly Poisson resum the metric component $g_{a\bar{a}}^{\rm inst}$ (\ref{gaainst1}) over the electric charges $n_e$,
obtaining at the leading order in $g_{\eff}^2$ expansion:
\begin{eqnarray}
\tilde{g}_{a\bar{a}}^{\rm inst}=\frac{8\pi}{g^4_{\rm eff}}\sum_{n\in \bbZ}\frac{|a|^2\cD(-i)}{M(n)^3}
\exp\left(-\frac{16\pi^2 R}{g_{\rm eff}^2}|M(n)|+i\theta_m+2 i n\Theta_{\rm eff}+i2\pi R \sum_{i=1}^{N_f} s_i F_i(n)\right)\,,\label{PSgaa}
\end{eqnarray}
where we have defined:
\begin{eqnarray}
M(n)&=&\sqrt{|a|^2+\left(\frac{\theta_e}{2\pi R}+\frac{n}{R}\right)^2}\,,\label{Mn}\\
F_i(n)&=& {\tm_i}+\left(\frac{\theta_e}{2\pi R}+\frac{n}{R}\right)\frac{|m_i|}{|a|}\,.\label{Fi}
\end{eqnarray}
Here $M(n)$ appearing in (\ref{PSgaa}) corresponds to the Euclidean action of the so-called ``twisted mono\-pole'' found in \cite{LeeYi}, which can be generated by applying large gauge transformation on the monopole compactified on $\mathbb{R}^3\times S^1$.
In taking the $2\pi R \to 0$ limit, all $M(n)$ diverge except $n=0$, therefore only the $n=0$ term survives in the summation. Furthermore, as $\theta_e/(2\pi R)$ is also kept fixed in such limit, $2\pi R F_i(0)$ tends to zero.
By again applying the three dimensional $SU(2)_N$ rotation symmetry,
we can deduce that the Poisson resummed metric (\ref{PSgaa}) corresponds to the following four fermion vertex in the three-dimensional effective Lagrangian:
\begin{equation}\label{3dS4F}
S_{4F}=\frac{2^{11}\pi^3 M_W R_H} {e_{\rm eff}^8}\exp\left(-\frac{4\pi}{e_{\rm eff}^2}M_W+i\theta_m\right)
\int d^3 x (\psi\cdot\bar{\psi})(\lambda\cdot\bar{\lambda})\,.
\end{equation}
Here we have kept the combination $1/e_{\rm eff}^2=2\pi R/g^2_{\rm eff}$ fixed, $R_H=\cR_q \cR_{\tq}$ is the hypermultiplet one-loop determinant computed in \cite{DTV}, and the W-boson mass is $M_W=2M(0)$.
This, after recalling the gauge coupling and electric charge for the W-boson, precisely matches with the four fermion vertex computed from first principles in \cite{Dorey1997} and \cite{DTV}.

\subsection*{Brane Picture}
\paragraph{}
It is possible to understand the form of many of the previous field
theory results in an elegant way in terms of Hanany-Witten brane
configurations \cite{HaWi}. In order to make the discussion more
transparent we will work in terms of a T-dual picture, in which
instead of a 4d theory compactified on a circle, we have a 3d field
theory localized in a compact transverse direction. Consider IIB
theory in the presence of two D3 branes with world volume coordinates
$(x^0 x^1 x^2 x^6)$ suspended between two NS5 branes with world volume
coordinates $(x^0 x^1 x^2 x^3 x^4 x^5)$ and sitting $L_6$ apart in the
$x^6$ direction. Additional $N_f$ D5 branes with world volume
coordinates $(x^0 x^1 x^2 x^7 x^8 x^9)$ provide the flavors.  The
Coulomb branch of the gauge theory in the $(x^0 x^1 x^2)$ directions
is realized when the two suspended D3 branes are split along $(x^3 x^4
x^5)$ with separation $\Delta\vec{x}$. We take $x^3$ to be the compact
direction in which we have T-dualized our original 4d theory; it has
dual radius $\tilde R=1/R$, with $R$ being the compactification radius for
the original 4d theory. We show the relevant brane configuration in
figure~1.
\begin{figure}[ht]
\label{fig:hanany-witten}
\centering
\includegraphics[width=0.5\textwidth]{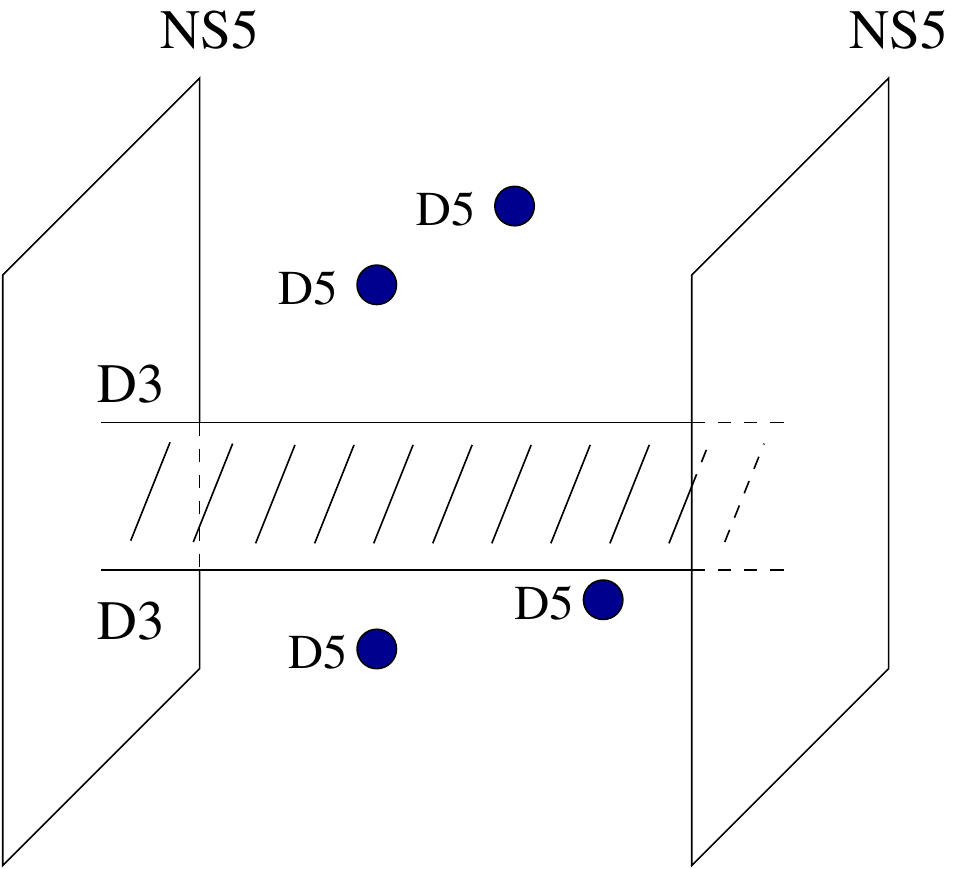}
\caption{Hanany-Witten configuration for the 3d $\cN=2$ $SU(2)$ with
  flavors theory described in the text. D1 branes (shaded) extending
  between the two D3 branes give rise to 3d instantons.}
\end{figure}
\paragraph{}
Non-perturbative instanton corrections to the Coulomb branch metric
come from Euclidean D1 strings stretching between D3s and NS5s, whose
world volume is bounded between the intervals $\Delta\vec{x}$ in $(x^3
x^4 x^5)$ and $L_6$ in $x^6$ \cite{dBHO, HaWi,dBHOOY}. The D1 DBI
action is given by the tension of the D1, it is $S_{\rm D1}\sim
-\frac{L_6 |\Delta\vec{x}(n)|}{g_s}$, where $|\Delta\vec{x}(n)|$ is
the norm of a vector $(\Delta x^3+2\pi n{\tilde{R}}=\Delta x^3+2\pi
n/R, \Delta{x}_4,\Delta{x}_5 )$. Due to the periodicity of $x^3$, we
need to sum over all multiply wound D1 branes, with winding number
given by $n\in \mathbb{Z}$.  Using the expression
$\frac{8\pi}{e^2_{\rm eff}}= \frac{L_6}{g_s}$ for the gauge coupling,
we see that $S_{\rm D1}$ coincides precisely with the real part of the
twisted monopole action in (\ref{PSgaa}). To account for the phase
$i\theta_m$ in (\ref{PSgaa}), which is the dual of unbroken $U(1)$
photon, we recall that the D1 action also receives a boundary
contribution, as D1s are charged magnetically under the D3 gauge
fields. This was made explicit in \cite{dBHO}, where the dual photon
was identified with the $x^6$ component of the magnetic gauge
potential $\tilde{A}_6$. Integrating over the boundary of D1s, we
obtain the desired phase.
\paragraph{}
The open string stretching between D3 sitting at $\vec{x}$ and the
additional D5 brane sitting at $\vec{m}_i$, where $\vec{x}$ and
$\vec{m}_i$, are their respective positions in the $(x^3,x^4,x^5)$
directions, gives fundamental quark of mass $|\vec{x}-\vec{m}_i|$. Again
due to the compactification in $x^3$ direction, it is necessary to
periodically identify $x^3-m^3_i \sim x^3-m^3_i +2\pi n/R$ and sum over
the copies. Combining with the W-boson mass $|\Delta \vec{x}(n)|$ and
appropriate weight for different representations, they explain the
form of perturbatively corrected gauge coupling (\ref{DefMn}), which
can alternatively be obtained from explicit one loop computation
following \cite{Dorey1997, SeiShe, DTV}.
\paragraph{}
Finally, the brane picture
also gives a geometrical understanding for the index computation
(\ref{DefIq}). There are D1-D5 strings and when they are localized at
the intersection points between D1 world volume and D5s, they become
additional hypermultiplet zero modes. Given that $(x^3 x^4 x^5)$
coordinates of D1 and D5 have natural interpretation of adjoint scalar
VEV/W-boson mass and quark bare mass, extra zero modes only appear
when their values coincide or both vanish. Specific combinations
depend on the choices of vacua, and in strict three dimensional limit,
they should be related by $SU(2)_N$ rotation. For specific example,
see \cite{dBHO}. 

\subsection*{Acknowledgements}
We would like to thank Nick Dorey for various useful discussions and comments on the draft.
HYC would like to thank Inaki Garcia-Etxebarria for inital collaboration and help with the figure,
he would also like to thank Gary Shiu and Peter Ouyang for discussions.
HYC is supported in part by NSF CAREER Award No. PHY-0348093, DOE grant
DE-FG-02-95ER40896, a Research Innovation Award and a Cottrell Scholar Award
from Research Corporation, and a Vilas Associate Award from the University of
Wisconsin.
KP is supported by a research studentship from Trinity College, Cambridge.

\end{document}